\documentclass[preprint,showpacs,preprintnumbers,amsmath,amssymb]{revtex4}

\pdfoutput=1

\usepackage{graphicx}
\usepackage{dcolumn}
\usepackage{bm}
\usepackage{amssymb}

\begin{document}
\title{
Gaussian wavepacket dynamics and quantum tunneling \\ 
in asymmetric double-well systems
}

\author{Hideo Hasegawa}
\altaffiliation{hideohasegawa@goo.jp}
\affiliation{Department of Physics, Tokyo Gakugei University,  
Koganei, Tokyo 184-8501, Japan}%

\date{\today}
\begin{abstract}
We have studied dynamical properties and quantum tunneling in asymmetric double-well (DW) systems,
by solving Schr\"{o}dinger equation with the use of two kinds of spectral methods
for initially squeezed Gaussian wavepackets.
Time dependences of wavefunction, averages of position
and momentum, the auto-correlation function, an uncertainty product and
the tunneling probability have been calculated. Our calculations have shown that
(i) the tunneling probability is considerably reduced by a potential asymmetry $\Delta U$,
(ii) a resonant tunneling with $\vert \Delta U \vert \simeq \kappa \:\hbar \omega$ 
is realized for motion starting from upper minimum 
of asymmetric potential wells, but not for motion from lower minimum 
($\kappa=0,1,2,\cdots$; $\omega$: oscillator frequency at minima), 
(iii) the reduction of the tunneling probability by an asymmetry is less significant
for the Gaussian wavepacket with narrower width, and
(iv) the uncertainty product $\langle \delta x^2 \rangle \langle \delta p^2 \rangle$ 
in the resonant tunneling state is larger than that in the non-resonant tunneling state.
The item (ii) is in contrast with the earlier study
[Mugnai {\it et al.}, Phys. Rev. A {\bf 38} (1987) 2182] which showed the symmetric result
for motion starting from upper and lower minima.

\vspace{1cm}


\end{abstract}

        

\maketitle
\newpage
\section{Introduction}
Double-well (DW) systems have been extensively studied in a wide range of fields
including physics, chemistry and biology (for a recent review on DW systems, see
Ref. \cite{Thorwart01}).
Quantum tunneling is one of the most fascinating phenomena in DW systems \cite{Razavy03}.
Much experimental and theoretical studies have been made
in tunneling of a quantum particle in DW systems.
Quantum tunneling of a particle is possible from one-side well 
to the other-side well through classically forbidden region.
Well-known old examples of DW systems include an inversion of anmonia molecule.
In recent years, there has been an advance in the experimental study
on macroscopic quantum tunneling such as Josephson junction
and Bose-Einstein condensation in a double trap.

DW potential does not have to be symmetric and it may be asymmetric in general.
In many experiments, the asymmetry of the DW potential can be changed 
by modifying external parameters.
However, most theoretical studies have been made for symmetric DW systems, 
and asymmetric systems have received less theoretical attention than symmetric ones
\cite{Weiner81,Nieto85,Mugnai88,Song08,Rastelli12,Cordes01}.
This is because solving an asymmetric DW system is more difficult than a symmetric one.
Theoretical studies on  asymmetric DW systems have been made based on 
various approximate methods like the WKB for simplified artificial DW potentials
which are analytically tractable but not realistic
\cite{Razavy03}.
By using such DW potentials, Weiner and Tse \cite{Weiner81}, 
and Nieto {\it et al.} \cite{Nieto85} showed that although the tunneling probability 
is significantly reduced by the potential asymmetry, it is enhanced when
the asymmetry meets the resonance condition.
Mugai {\it et al.} \cite{Mugnai88} studied the fractal nature of the trajectory 
in asymmetric DW systems.
By using WKB, Song \cite{Song08} studied an asymmetric DW system 
where the difference of the potential minima is close of a multiple of $\hbar \omega$
(harmonic frequency in the wells).
Rastelli \cite{Rastelli12} obtained a semi-classical formula for the tunneling amplitude 
in asymmetric DW systems with the use of WKB method.
Conventional theories for DW systems have adopted the two-level approximation where
the initial state in one-dimensional system is assumed to be given by
$\Psi(x,0)=[\Psi_0(x)-\Psi_1(x)]/\sqrt{2}$, $\Psi_{\nu}(x)$ denoting the 
$\nu$th ($\nu=0, 1$) eigenfunction. 
In order to discuss the tunneling probability in asymmetric DW systems,
Cordes and Das \cite{Cordes01} proposed a generalized two-level approximation:
related discussion will be given in Sec. IV.

For a study on dynamics of wavepacket or tunneling in DW systems,
it is necessary to solve the time-dependent Sch\"{o}dinger equation
subject to appropriate initial and boundary conditions \cite{Tannor07}.
In the past when quantum mechanics was born, it was very difficult
to numerically solve the time-dependent Sch\"{o}dinger equation 
even for a simple potential except for a harmonic oscillator (HO) potential.
One had to develop approximation methods applicable to simple tractable DW models 
although they are not necessarily realistic.
In recent years, however, there has been significant development
in computer and its software. It is now possible for us
to solve the time-dependent Schr\"{o}dinger equation with sufficient accuracy,
by using convenient packages such as MATHEMATICA, MATLAB and Maple.

The purpose of the present study is to numerically study dynamics of
Gaussian wavepackets and to examine the effect of the asymmetry
on quantum tunneling in asymmetric DW systems.
Quite recently it has been pointed out that a potential asymmetry of a DW system 
has significant effects on its specific heat \cite{Hasegawa12b}.
We expect that it is the case also for dynamical properties of DW systems.
We will solve the time-dependent Sch\"{o}dinger equation by the spectral method
for a given squeezed Gaussian wavepacket \cite{Cooper86,Tsue91}, 
adopting the realistic quartic DW potential.
In order to investigate the influence of the initial state on dynamical properties,
we adopt two squeezed Gaussian wavepackets with different parameters.

The paper is organized as follows.
In Sec. II, we will briefly mention the model and calculation method employed 
in our study \cite{Hasegawa13a}.
In solving the time-dependent Schr\"{o}dinger equation, we have adopted the two kinds of
spectral method A [Eq. (\ref{eq:B7})] and spectral method B [Eq. (\ref{eq:C4})] 
with energy matrix elements evaluated for a finite size $N_m$ ($=30$).
By using the spectral method A, we have calculated time-dependences of 
the magnitude of wavefunction, expectation values of position and momentum, 
the auto-correlation function, the uncertainty product 
and the tunneling probability, whose results are reported in Sec. III.
In Sec. IV the tunneling probability is discussed with the use of the spectral method B.
We discuss also wavepacket dynamics when the Gaussian wavepacket
starts from near the top of the DW potential. 
Sec. V is devoted to our conclusion.

\section{Adopted model and calculation method}
\subsection{Asymmetric double-well systems}
\begin{figure}
\begin{center}
\includegraphics[keepaspectratio=true,width=70mm]{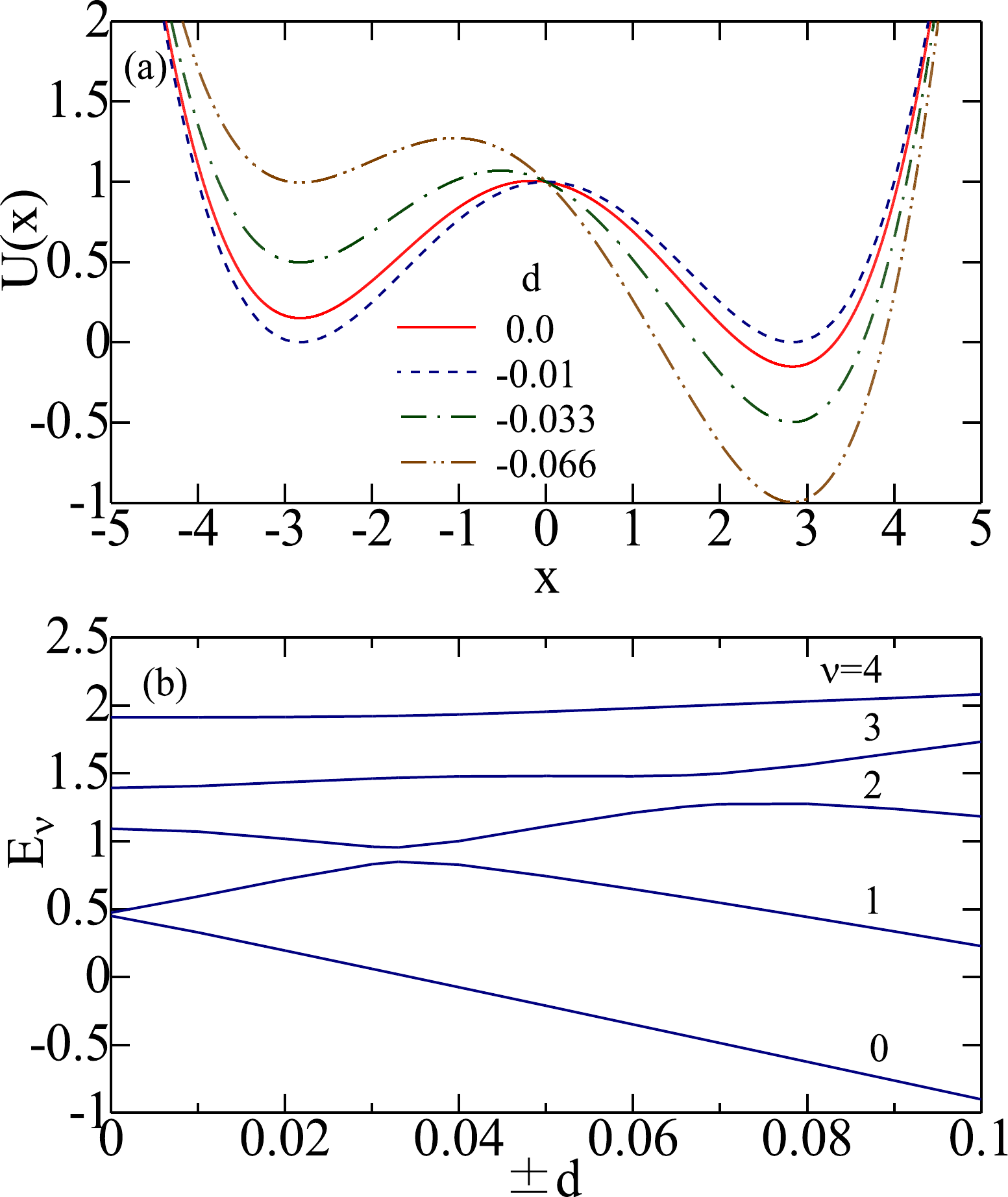}
\end{center}
\caption{
(Color online) 
(a) The asymmetric DW potential for $d=0.0$ (solid curve), $d=-0.01$ (dashed curve),
$d=-0.033$ (chain curve) and $d=-0.066$ (double-chain curve) 
with $x_s=2\sqrt{2}$ and $m=\omega=1.0$ in Eq. (\ref{eq:L1}).
(b) Eigenvalues of $E_{\nu}$ ($\nu=0$ to $4$) as a function of $\pm d$.
$E_{\nu}$ and $U(x)$ are symmetric and anti-symmetric, respectively, 
with respect to a sign of $d$.
}
\label{fig1}
\end{figure}

We assume a quantum DW system whose Hamiltonian is given by 
\begin{eqnarray}
H &=& \frac{p^2}{2 m} + U(x) = H_0+V(x),
\label{eq:B1}
\end{eqnarray}
where
\begin{eqnarray}
U(x) &=& C\;(x^2-x_s^2)^2
-d \left( \frac{x^3}{3}-x_s^2 x\right),
\hspace{0.5cm}\mbox{$\left( C=\frac{m \omega^2}{8 x_s^2} \right)$} 
\label{eq:L1} \\
H_0 &=& \frac{p^2}{2m}+U_0(x), \\
U_0(x) &=& \frac{m \omega^2 x^2}{2}, \\
\label{eq:H2}
V(x) &=& U(x)-U_0(x). 
\label{eq:B4}
\end{eqnarray}
Here $m$, $x$ and $p$ express mass, position and momentum, respectively, of a particle,
$U(x)$ denotes the DW potential with a degree of the asymmetry $d$, 
$H_0$ signifies the Hamiltonian 
for an HO potential $U_0(x)$ with the oscillator frequency $\omega$,
and $V(x)$ stands for a perturbing potential to $H_0$.
The asymmetric DW potential $U(x)$ has locally stable minima at $x=\pm x_s$ 
and an unstable maximum at $x_u=d(2 x_s^2/m \omega^2)$ with 
\begin{eqnarray}
U(\pm x_s) &=& \pm \frac{2 d x_s^3}{3}, \\
U(x_u) &=& \frac{m \omega^2 x_s^2}{8}+\frac{d^2 x_s^2}{m \omega^2}
-\frac{2 d^4 x_s^4}{3 m^3 \omega^6}, \\
\Delta U &=& U(x_s)-U(-x_s)=\frac{4 d x_s^3}{3}.
\end{eqnarray}
A prefactor of $C$ $(=m \omega^2/8 x_s^2)$ in Eq. (\ref{eq:L1}) is chosen such that
the DW potential $U(x)$ for $d=0.0$ has the same curvature at the minima 
as the HO potential $U_0(x)$: $U''(\pm x_s)=U_0''(0)=m \omega^2$ 
\cite{Hasegawa12b,Hasegawa13a}.
The asymmetry parameter $d$ is assumed to be given by
\begin{eqnarray}
-d_c < d < d_c = \frac{m \omega^2}{2 x_s},
\end{eqnarray}
for which $x_u$ locates at $-x_s < x_u < x_s$.
In our model calculations, 
we have adopted parameters of $m=\omega=1.0$ and $x_s=2 \sqrt{2}$ 
which yield $d_c=0.1768$ and $U''(\pm x_s)=U_0''(0)=1.0$ for $d=0.0$.
The DW potential given by Eq. (\ref{eq:L1}) for typical values of $d=0.0$ (solid curve), 
$d=-0.01$ (dashed curve) and $d=-0.033$ (chain curve)
is plotted in Fig. \ref{fig1}(a). 

For the HO Hamiltonian $H_0$,
eigenfunction $E_{0n}$ and eigenvalue $\phi_n(x)$ are given by
\begin{eqnarray}
\phi_n(x) &=& \frac{1}{\sqrt{2^n n!}} 
\left( \frac{m \omega}{\pi \hbar} \right)^{1/4}
\exp\left( -\frac{m \omega x^2}{2 \hbar}\right)
H_n\left( \sqrt{\frac{m \omega}{\hbar}}\:x \right), 
\label{eq:A12}\\
E_{0n} &=& \left( n+\frac{1}{2} \right) \hbar \omega
\hspace{1cm}\mbox{($n=0,1,2\cdot,\cdot\cdot\cdot$)},
\label{eq:H3}
\end{eqnarray}
where $H_n(x)$ stands for the Hermite polynomials.

For the stationary state, we solve the time-independent Schr\"{o}dinger equation, 
expanding the eigenfunction $\Psi(x)$ in terms of $\phi_n(x)$
\begin{eqnarray}
\Psi(x) &=& \sum_{n=0}^{N_m} c_n \phi_n(x),
\label{eq:B5}
\end{eqnarray}
leading to the secular equation
\begin{eqnarray}
E \:c_n &=& \sum_{k=0}^{N_m} H_{n k} \:c_k,
\hspace{1cm}\mbox{($n=0$ to $N_m$)}
\label{eq:B6}
\end{eqnarray}
where $E$ denotes the eigenvalue and $N_m$ is the maximum quantum number.
From a diagonalization of the secular equation, we obtain the eigenvalue $E_{\nu}$
and its relevant eigenfunction $\Psi_{\nu}(x)$ 
satisfying
\begin{eqnarray}
H \Psi_{\nu}(x) &=& E_{\nu} \Psi_{\nu}(x).
\hspace{1cm}\mbox{($\nu=0$ to $N_m$)}
\label{eq:C5}
\end{eqnarray}

Figure \ref{fig1}(b) shows eigenvalues $E_{\nu}$ with $\hbar=1.0$
for $\nu=0-4$ as a function of $\pm d$.
Table 1 shows $U(\pm x_s)$, $U(x_u)$, $\Delta U$ ($=U(x_s)-U(-x_s)$), $\delta$ ($=E_1-E_0$)
and $\delta'$ ($=E_2-E_1$) as a function of the asymmetry $d$.
For $d > 0$ and $d < 0$, $\Delta U$ become $\Delta U > 0$ and $\Delta U < 0$, respectively.
With increasing $\vert d \vert$, both $\vert \Delta U \vert$ and 
$\delta$ are increased.
For $\vert d \vert \geq 0.02$, $\delta'$  becomes smaller than $\delta$.

\begin{center}
\begin{tabular}[b]{|c|c|c|c|c|c|c|}
\hline
$d$ & $U(-x_s)$ & $U(x_u)$ & $U(x_s)$ &  $\Delta U$ & $\delta$ & $\delta'$
\\ \hline \hline
0.0 & 0.0  & $\;\;1.0000 \;\;$ &  0.0 &  0.0 & $\;\;0.023923\;\;$ &\;\; 0.61849 \;\;\\
$\pm 0.010$ & $\mp 0.150849$ & 1.0064 &  $\pm 0.150849$ &  $\pm 0.301699$ & $0.264823$ & 0.47753  \\
$\pm 0.020$ & $\mp 0.301699$ & 1.02555 &  $\pm 0.301699 $ &  $\pm 0.603398$ & $0.525453$ & 0.29749 \\
%
%
$\pm 0.033$ & $\mp 0.497803$ & 1.06929 &  $\pm 0.497803 $ &  $\pm 0.995606$ & $0.829368$ & 0.10581  \\
$\pm 0.040$ & $\mp 0.603398$ & 1.10153 &  $\pm 0.603398 $ &  $\pm 1.2068$ & $0.902893$ & 0.17417  \\
$\pm 0.050$ & $\mp 0.754247$ & 1.15787 &  $\pm 0.754247 $ &  $\pm 1.50849$ & $0.955426$ & 0.36639  \\
$\pm 0.066$ & $\mp 0.995606$ & 1.27231 &  $\pm 0.995606 $ &  $\pm 1.99121$ & $1.01905$ & 0.66749  \\
\hline
\end{tabular}
\end{center}

{\it Table 1} Potential values at locally-stable minima ($\pm x_s$) 
and an unstable maximum position ($x_u$), $\Delta U$ [$=U(x_s)-U(-x_s)$],
and energy gaps ($\delta=E_1-E_0$, $\delta' =E_2-E_1$)  
as a function of the asymmetry $d$ for the asymmetric DW potential 
[Eq. (\ref{eq:L1})] ($N_m=30$).

\subsection{Spectral method A}
For the non-stationary state, we solve the time-dependent Schr\"{o}dinger equation given by
\begin{eqnarray}
i \hbar \:\frac{\partial \Psi(x,t)}{\partial t} &=& H \:\Psi(x,t).
\label{eq:B7a}
\end{eqnarray}
In the spectral method A, the eigenfunction $\Psi(x,t)$ is expanded 
in terms of $\phi_n(x)$
\begin{eqnarray}
\Psi(x,t) &=& \sum_{n=0}^{N_m} c_n(t) \phi_n(x),
\label{eq:B7}
\end{eqnarray}
where $c_n(t)$ stands for the time-dependent expansion coefficient
obeying equations of motion given by
\begin{eqnarray}
i \hbar \:\frac{\partial c_n(t)}{\partial t} &=& \sum_{k=0}^{N_m} H_{nk} \:c_k(t)
\hspace{1cm}\mbox{($n=0$ to $N_m$)},
\label{eq:B8}
\end{eqnarray}
with
\begin{eqnarray}
H_{n k} &=& E_{0n}\:\delta_{n, k}
+\int_{-\infty}^{\infty} \phi_{n}(x)^* \:V(x) \:\phi_k(x)\;dx.
\label{eq:B9}
\end{eqnarray}
Equation (\ref{eq:B8}) expresses the ($N_m+1$) first-order differential equations, 
which may be solved for a given initial condition of $\{ c_n(0) \}$.
An initial value of the expansion coefficient $c_n(0)$ is determined by 
\begin{eqnarray}
c_n(0) &=& \int_{-\infty}^{\infty} \phi_n(x)^* \:\Psi_G(x, 0)\:dx,
\label{eq:B10}
\end{eqnarray}
for the squeezed coherent Gaussian wavepacket $\Psi_G(x, 0)$ expressed by
 \cite{Cooper86,Tsue91}
\begin{eqnarray}
\Psi_G(x, 0)=\frac{1}{(2 \pi \mu)^{1/4}} \:\exp\left[-\frac{(1-i \alpha)}{4 \mu}(x-x_0)^2
+ i \:\frac{p_0 (x-x_0)}{\hbar} \right],
\label{eq:B11}
\end{eqnarray}
where $x_0$ and $p_0$ are initial position and momentum, respectively, and
parameters $\mu$ and $\alpha$ are related with
\begin{eqnarray}
\langle \delta x^2\rangle &=& \mu, \;\;\;
\langle \delta x \delta p+ \delta p \delta x \rangle = \alpha.
\end{eqnarray}

\subsection{Spectral method B}
In an alternative spectral method B, the solution of the time-dependent Schr\"{o}dinger 
equation given by Eq. (\ref{eq:B7a}) is expressed by
\begin{eqnarray}
\Psi(x,t) &=& \sum_{\nu=0}^{N_m} a_{\nu} \:\Psi_{\nu}(x) \:e^{-i E_{\nu} t/\hbar},
\label{eq:C4}
\end{eqnarray}
where $\Psi_{\nu}(x)$ and $E_{\nu}$ are eigenfunction and eigenvalue of
the stationary state given by Eq. (\ref{eq:C5}).
Note that the expansion coefficient $a_{\nu}$ in Eq. (\ref{eq:C4}) is time independent
and it is determined by a given Gaussian wavepacket 
\begin{eqnarray}
a_{\nu} &=& \int_{-\infty}^{\infty} \Psi_{\nu}(x)^* \:\Psi_G(x, 0)\:dx.
\label{eq:C6}
\end{eqnarray}

Both spectral methods A and B yield the same result. 
Calculations of 
various time-dependent averages obtained by the spectral method A, 
which are presented in the Appendix,
are easier than those by the spectral method B, 
while the latter method is physically more transparent than the former.
By using mostly the spectral method A, we have performed model calculations 
to be reported in Sec. III. 
The spectral method B is employed for a discussion on the tunneling probability in Sec. IV.
Wavefunctions obtained by the spectral methods have been cross-checked by the 
MATHEMATICA resolver for the partial differential equation.

\section{Model calculations}

By using the method described in the preceding section, we have studied dynamics of 
Gaussian wavepackets in DW systems.  
Matrix elements $H_{n k}$ in Eq. (\ref{eq:B9}) for the adopted DW potential 
are given by Eq. (\ref{eq:H6}) in the Appendix. 
Model calculations for symmetric and asymmetric cases will
be separately reported in Secs. III A and III B, respectively.

\begin{figure}
\begin{center}
\includegraphics[keepaspectratio=true,width=70mm]{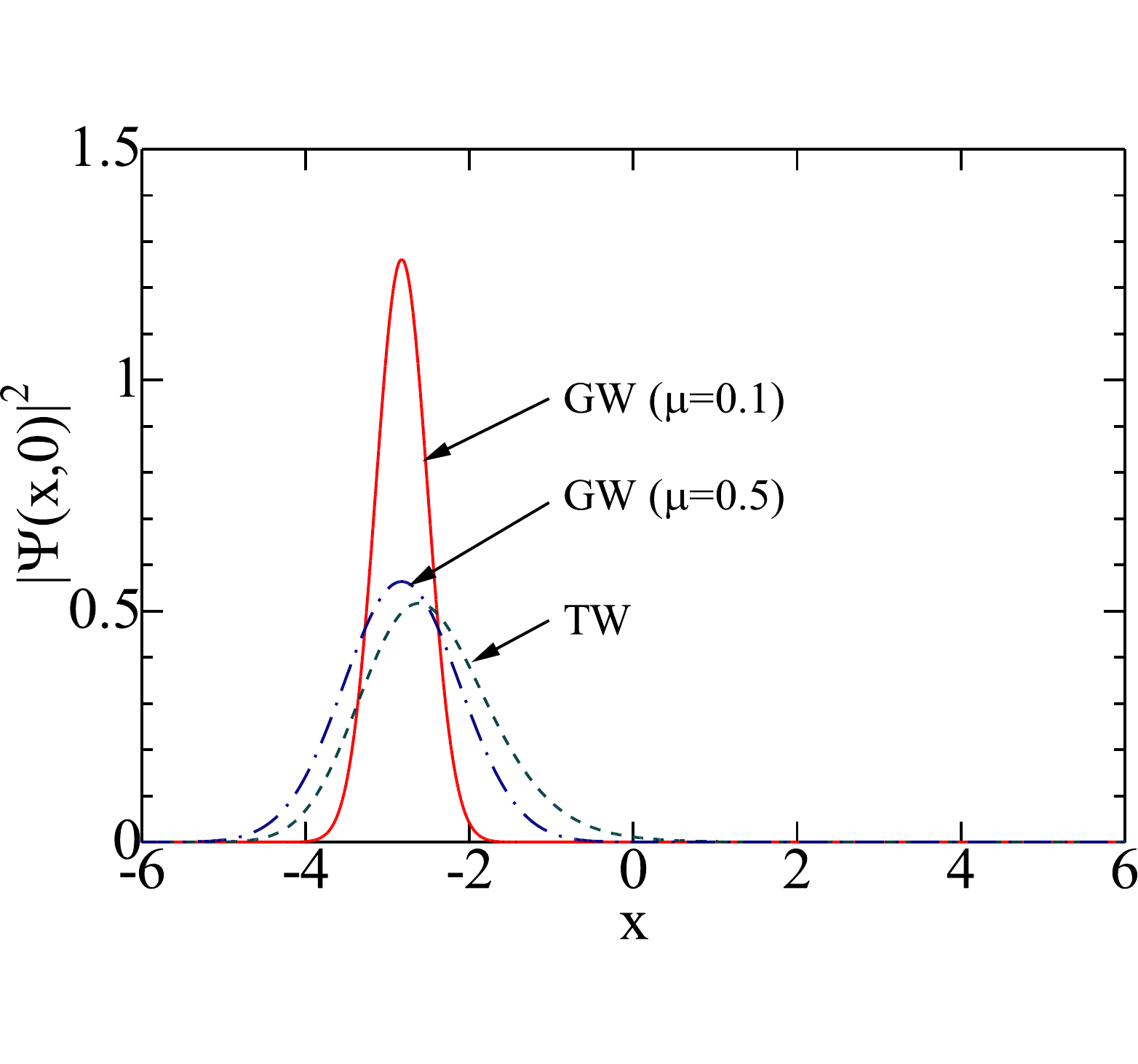}
\end{center}
\caption{
(Color online) 
Magnitudes of Gaussian wavepackets (GWs) with $\mu=0.1$ (solid curve) 
and $\mu=0.5$ (chain curve), dashed curve showing the two-level wavepacket (TW):
$\Psi(x,0)=[\Psi_0(x)-\Psi_1(x)]/\sqrt{2}$. 
}
\label{fig2}
\end{figure}

\begin{figure}
\begin{center}
\includegraphics[keepaspectratio=true,width=70mm]{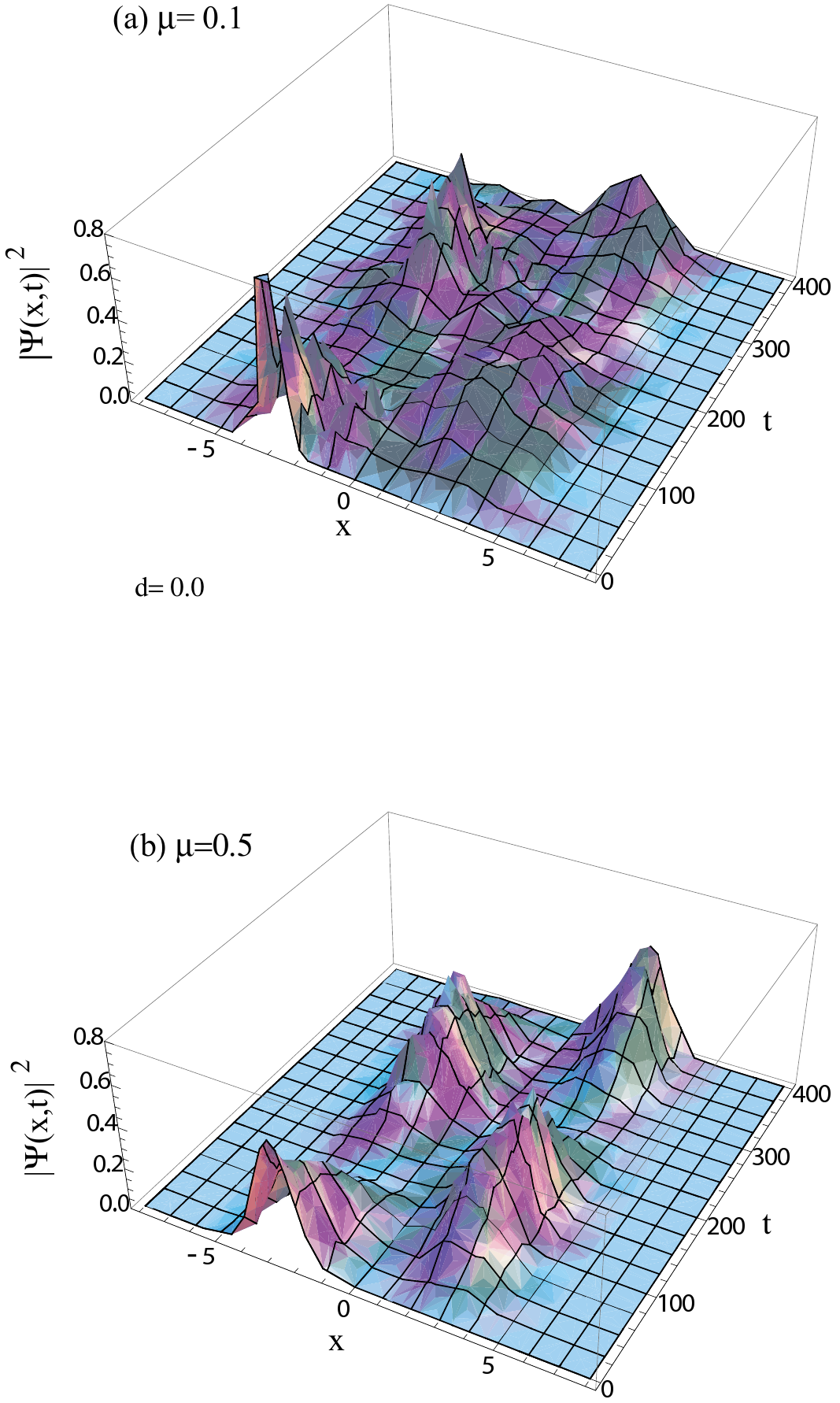}
\end{center}
\caption{
(Color online) 
3D plots of $\vert \Psi(x,t) \vert^2$ as functions of $x$ and $t$ for  
Gaussian wavepackets with (a) $\mu=0.1$ and (b) $\mu=0.5$ in the symmetric DW system 
($x_0=-2 \sqrt{2}, p_0=0.0$). 
}
\label{fig3}
\end{figure}

\begin{figure}
\begin{center}
\includegraphics[keepaspectratio=true,width=70mm]{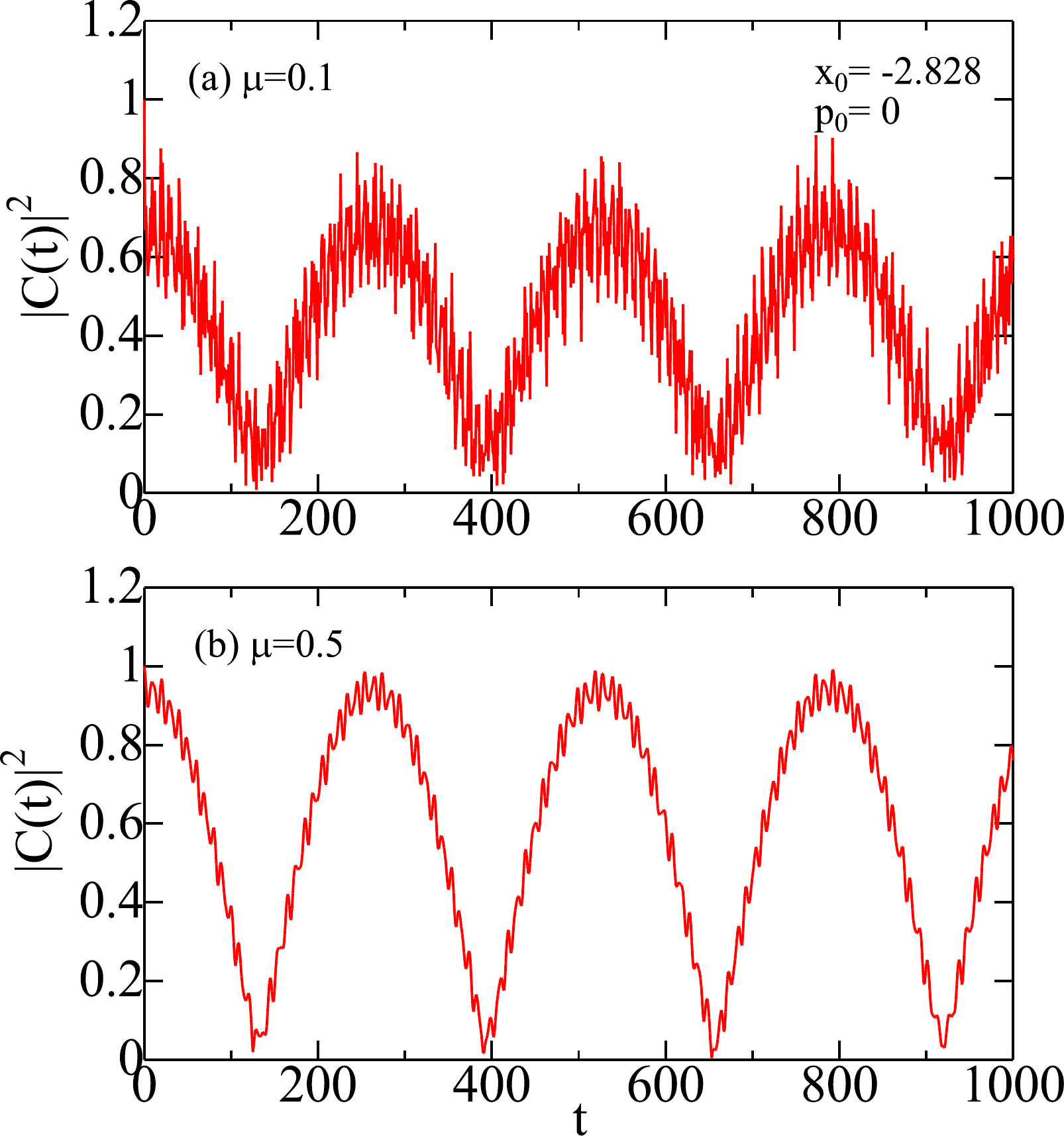}
\end{center}
\caption{
(Color online) 
The time dependence of the auto-correlation function $\vert C(t) \vert^2$
in the symmetric DW system calculated for Gaussian wavepackets 
with (a) $\mu=0.1$ and (b) $\mu=0.5$ ($x_0=-2\sqrt{2}, p_0=0.0$).
}
\label{fig4}
\end{figure}

\begin{figure}
\begin{center}
\includegraphics[keepaspectratio=true,width=70mm]{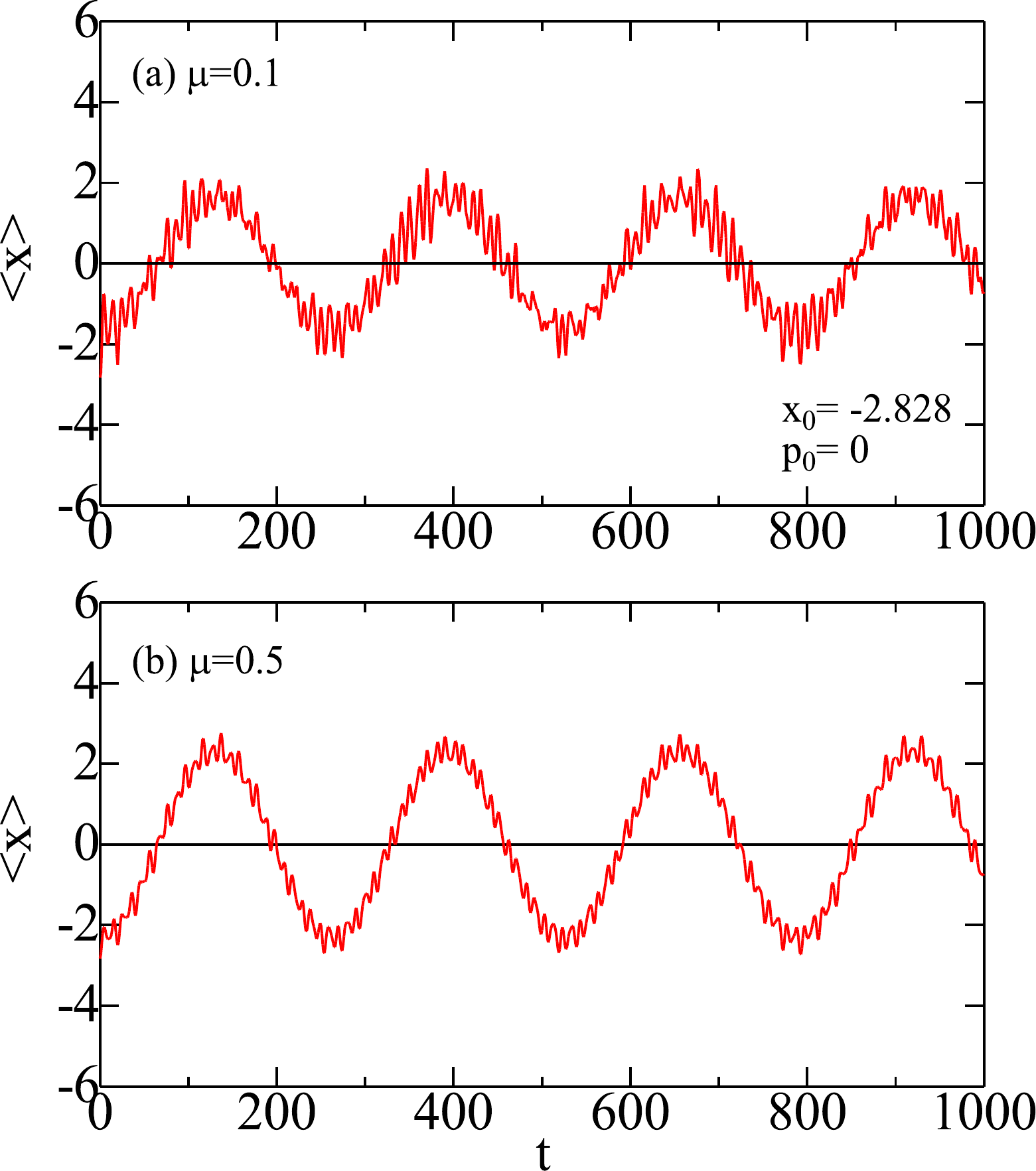}
\end{center}
\caption{
(Color online) 
The time dependence of $\langle x \rangle$ in the symmetric DW system calculated 
for Gaussian wavepackets with (a) $\mu=0.1$ and (b) $\mu=0.5$ ($x_0=-2\sqrt{2}, p_0=0.0$).
}
\label{fig5}
\end{figure}

\begin{figure}
\begin{center}
\includegraphics[keepaspectratio=true,width=70mm]{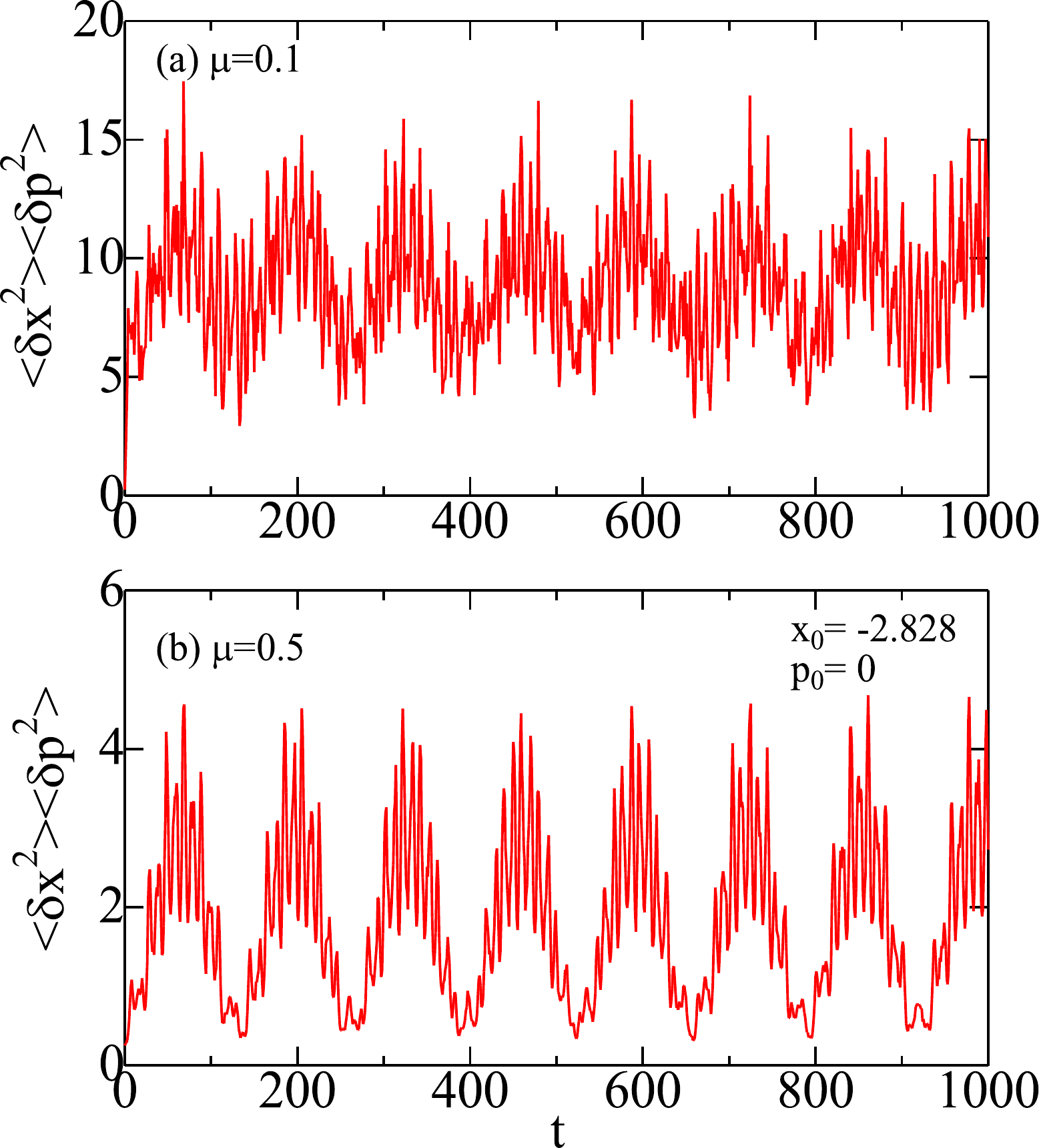}
\end{center}
\caption{
(Color online) 
The time dependence of the uncertainty product: 
$\langle \delta x^2 \rangle \langle \delta p^2 \rangle$ in the symmetric DW system calculated 
for Gaussian wavepackets with (a) $\mu=0.1$ and (b) $\mu=0.5$ ($x_0=-2\sqrt{2}, p_0=0.0$).
}
\label{fig6}
\end{figure}

\subsection{Symmetric case}
First we consider the case of the symmetric potential with $d=0.0$.
A diagonalization of the energy matrix with $N_m=30$ leads to
eigenvalues of $E_{\nu}=$ 0.450203, 0.474126, 1.09262, 1.39334 and 1.91286
for $\nu=0$ to 4, respectively, which are plotted in Fg. \ref{fig1}(b). 
The ground and first-excited states are quasi-degenerate with a energy gap
of $\delta = 0.023923$.
In order to examine effects of the Gaussian wavepacket on dynamical properties,
we consider the two Gaussian wavepackets given by Eq. (\ref{eq:B11})
with $\mu=0.1$ and $\mu=0.5$ for $x_0=-x_s$, $p_0=0.0$ and $\alpha=0.0$,
which are plotted by solid and chain curves, respectively, in Fig. \ref{fig2}.
The dashed curve will be explained later (Sec. IV).

Figures \ref{fig3}(a) and \ref{fig3}(b) show 3D plots of $\vert \Psi(x,t) \vert^2$
calculated by  Gaussian wavepackets with $\mu=0.1$ and $\mu=0.5$, respectively.
As time is developing, initial Gaussian wavepackets are deformed, 
and wavepackets at $t > 0$ cannot be expressed by a single Gaussian \cite{Hasegawa13a}. 

Figures \ref{fig4}(a) and \ref{fig4}(b) show the auto-correlation function
$\vert C(t) \vert^2$ for the Gaussian wavepackets with $\mu=0.1$ and $\mu=0.5$, respectively.
Both auto-correlation functions oscillate with a period of about 260, which is consistent
with the period given by $T=2 \pi/\delta=262$.
 
Figures \ref{fig5}(a) and \ref{fig5}(b) show expectation values of $\langle x \rangle$
for Gaussian wavepackets with $\mu=0.1$ and $\mu=0.5$, respectively.
It is clearly seen that a particle tunnels between the left and right wells
with a period of about 260.

Figures \ref{fig6}(a) and \ref{fig6}(b) show the uncertainty product,
$\langle \delta x^2 \rangle \langle \delta p^2 \rangle$, 
for Gaussian wavepackets with $\mu=0.1$ and $\mu=0.5$, respectively,
where $\delta x=x- \langle x \rangle$ and $\delta p=p- \langle p \rangle$.
They start from the minimum uncertainty of $\hbar^2/4$ at $t=0$ and oscillate
with fairly large magnitudes and
with a period of about 130, a half of the period of $\langle x \rangle$ in Fig. \ref{fig5}.
Its magnitude for $\mu=0.1$ is larger than that for $\mu=0.5$
by a factor of about four. 

\subsection{Asymmetric case}

\begin{figure}
\begin{center}
\includegraphics[keepaspectratio=true,width=80mm]{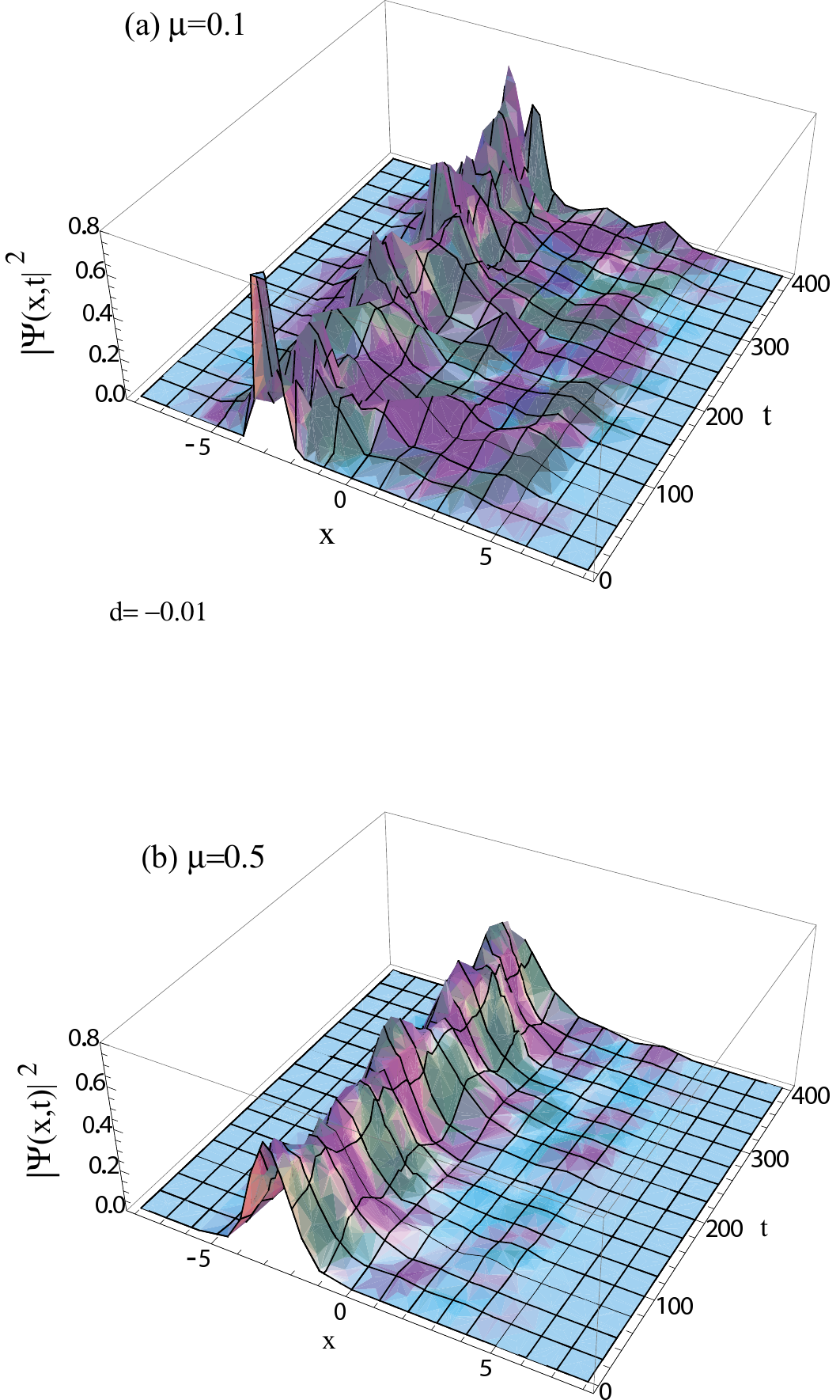}
\end{center}
\caption{
(Color online) 
3D plots of $\vert \Psi(x,t) \vert^2$ as functions of $x$ and $t$ 
calculated by Gaussian wavepackets with (a) $\mu=0.1$ and (b) $\mu=0.5$ 
in the asymmetric DW system with $d=-0.01$ ($x_0=-2 \sqrt{2}, p_0=0.0$). 
}
\label{fig7}
\end{figure}

\begin{figure}
\begin{center}
\includegraphics[keepaspectratio=true,width=120mm]{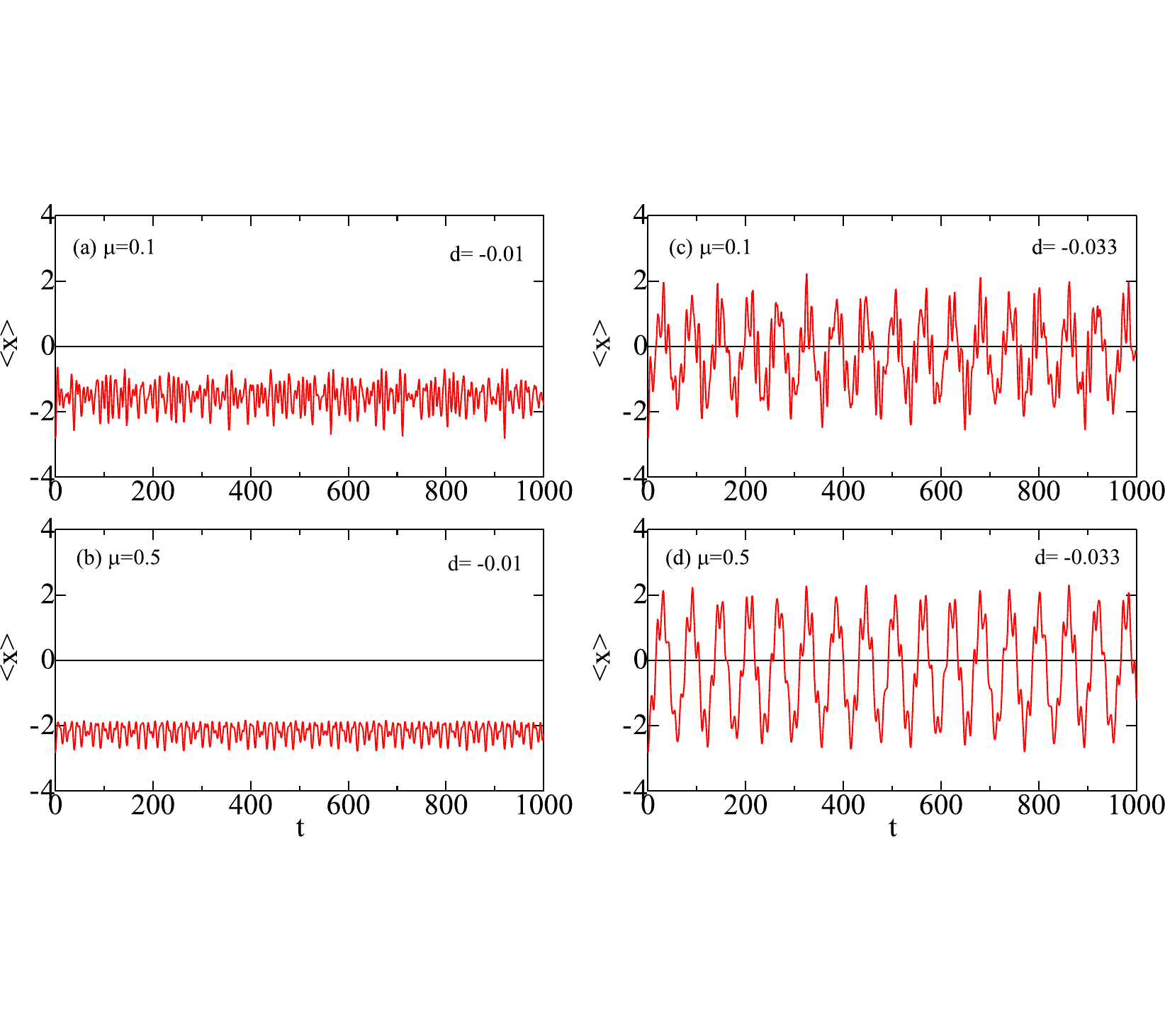}
\end{center}
\caption{
(Color online) 
The time dependence of $\langle x \rangle$ in the asymmetric DW system 
calculated by Gaussian wavepackets with (a) $\mu=0.1$ and (b) $\mu=0.5$ 
for $d=-0.01$, and with (c) $\mu=0.1$ and (d) $\mu=0.5$ for $d=-0.033$.
}
\label{fig8}
\end{figure}

Next we consider the asymmetric case with $d \neq 0$.
For $d=-0.01$, the potential minimum in the right well is lower than
that in the left well by $\Delta U=-0.301699$ (Table 1).
We obtain eigenvalues of $E_{\nu}=$
0.328786, 0.59361, 1.07114, 1.40643 and 1.91312 for $\nu=0-4$, respectively,
which are plotted in Fig. \ref{fig1}(b).
Quasi-degeneracy between $E_0$ and $E_1$ for $d=0.0$ is removed
by an introduced asymmetry, while $E_2$, $E_3$ and $E_4$ are almost
independent of $d$.

Figures \ref{fig7}(a) and \ref{fig7}(b) show 3D plots of $\vert \Psi(x,t) \vert^2$
for $d=-0.01$ calculated by Gaussian wavepackets with $\mu=0.1$ and $\mu=0.5$, respectively,
for $x_0=-2\sqrt{2}$, $p_0=0.0$ and $\alpha=0.0$.
A comparison between Fig. \ref{fig7}(a) [Fig. \ref{fig7}(b)] 
and Fig. \ref{fig3}(a) [Fig. \ref{fig3}(b)]
shows that $\vert \Psi(x,t) \vert^2$ for $d=-0.01$ stays in the left well
and tunneling of a particle is almost vanishing.
This is more clearly seen in Figs. \ref{fig8}(a) and \ref{fig8}(b) 
which show time dependences of $\langle x \rangle$ for $\mu=0.1$ and $\mu=0.5$, respectively..

We furthermore increase the asymmetry to $d=-0.033$, for which 
eigenvalues are $E_{\nu}=$
0.0193182, 0.848686, 0.954496, 1.46802 and 1.92407, respectively [Fig. \ref{fig1}(b)].
The energy gap of $\delta\;(=E_1-E_0)=0.829368$ is larger than
$\delta'\;(=E_2-E_1)=0.10581$, and
the difference between two potential minima becomes $\Delta U = -0.995606$ (Table 1).
We note that $\Delta U \simeq \omega=1.0$, for which a resonance of tunneling is expected.
Indeed, expectation values of $\langle x \rangle$ for $\mu=0.1$ (Fig. \ref{fig8}(c))
and $\mu=0.5$ (Fig. \ref{fig8}(d))
show tunneling with a period of about 60. This figure agrees with
$2 \pi/\delta'=59.382$, which implies that contributions from the first- and second-excited
states play important roles in the case of $d=-0.033$.

\begin{figure}
\begin{center}
\includegraphics[keepaspectratio=true,width=120mm]{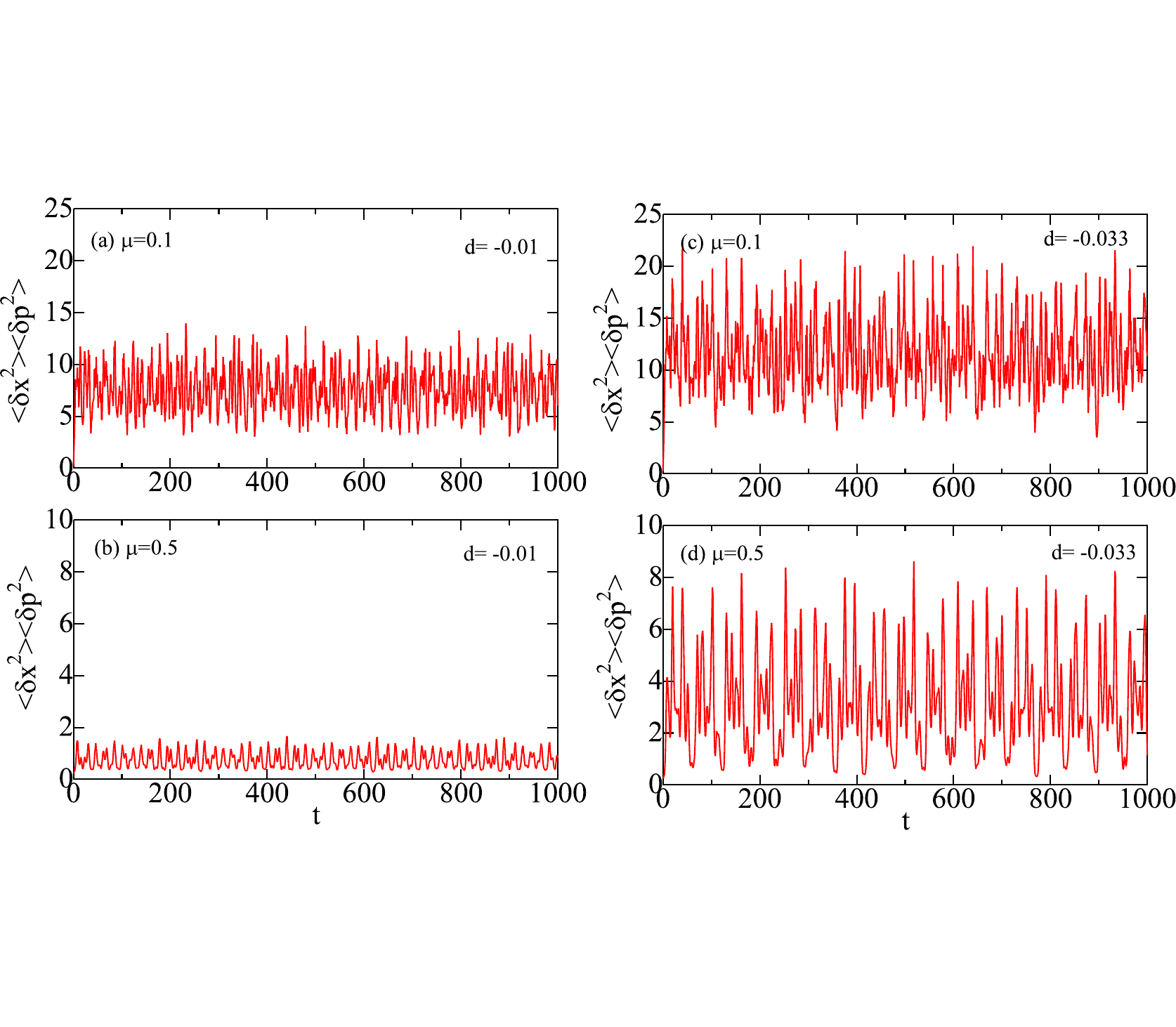}
\end{center}
\caption{
(Color online) 
The time dependence of $\langle \delta x^2 \rangle \langle \delta p^2 \rangle$ 
in the asymmetric DW system 
calculated by Gaussian wavepackets with (a) $\mu=0.1$ and (b) $\mu=0.5$ for $d=-0.01$,
and with (c) $\mu=0.1$ and (d) $\mu=0.5$ for $d=-0.033$.
}
\label{fig9}
\end{figure}

Figures \ref{fig9}(a)-\ref{fig9}(d) show time dependences of the uncertainty product
of $\langle \delta x^2 \rangle \langle \delta p^2 \rangle$ for $d=-0.01$ and $-0.033$,
which should be compared to those for $d=0.0$ shown in Fig. \ref{fig6}.
The uncertainty product for $d=-0.01$ is smaller than that for $d=0.0$.
It is, however, again increased for $d=-0.033$. We note that
$\langle \delta x^2 \rangle \langle \delta p^2 \rangle$ 
in the resonant tunneling state with $d=0.0$ or $d=-0.033$ 
is larger than that in the non-resonant tunneling state with $d=-0.01$. 
This is mainly due to the fact that $\langle \delta x^2 \rangle$
in the former state is larger than that in the latter.
Magnitudes of uncertainty product for the Gaussian wavepacket with $\mu=0.1$ 
are larger than that with $\mu=0.5$. 

\begin{figure}
\begin{center}
\includegraphics[keepaspectratio=true,width=70mm]{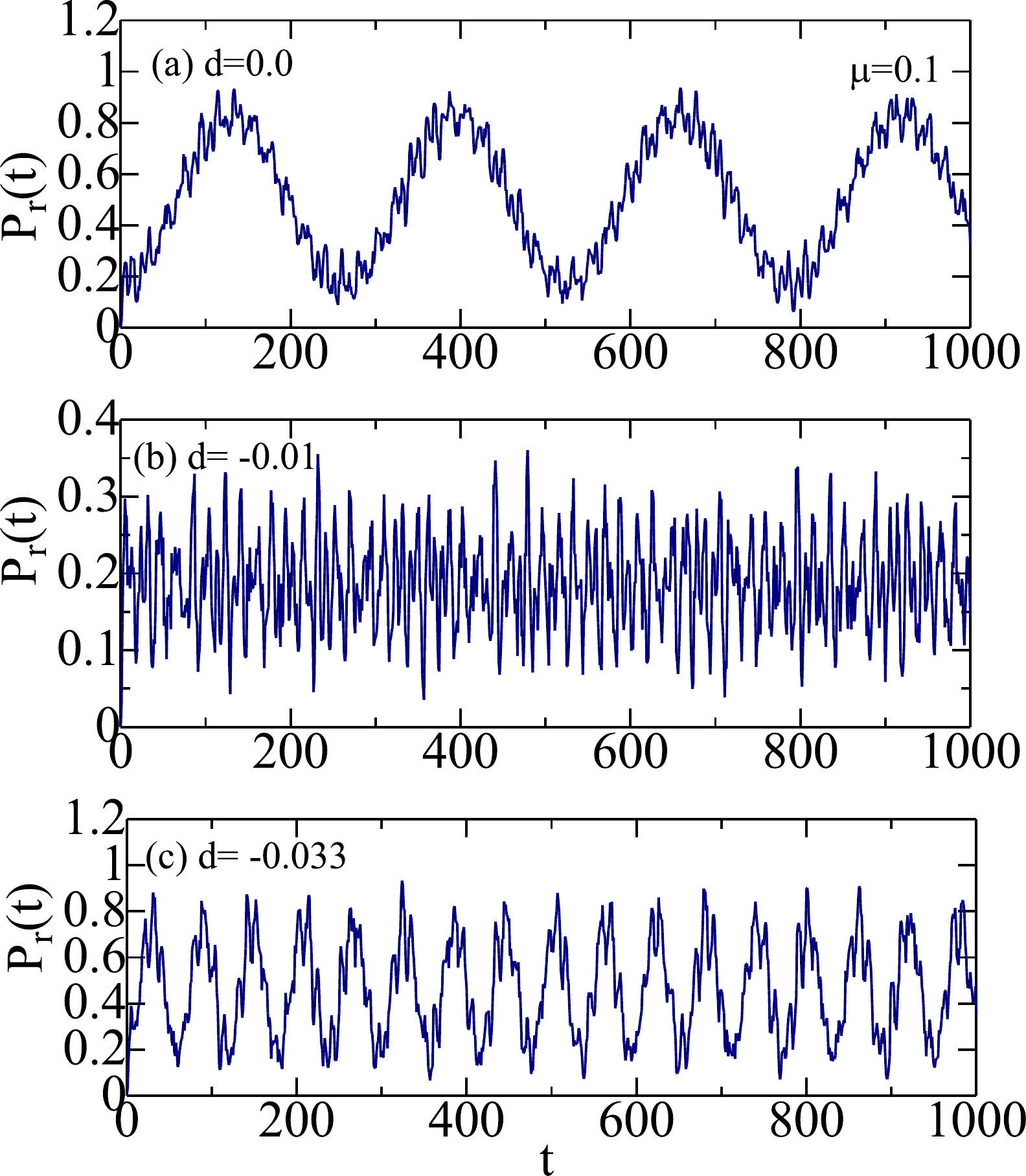}
\end{center}
\caption{
(Color online) 
The time dependence of $P_r(t)$ in the asymmetric DW system 
with (a) $d=0.0$, (b) $d=-0.01$ and (c) $d=-0.033$
calculated by the Gaussian wavepacket with $\mu=0.1$.
}
\label{fig10}
\end{figure}

The tunneling probability of $P_{r}(t)$ for finding a particle in the right well 
is defined by
\begin{eqnarray}
P_{r}(t) &=& \int_0^{\infty} \Psi(x,t)^* \Psi(x,t)\:dx,
\label{eq:D1}
\end{eqnarray}
and its maximum by
\begin{eqnarray}
P_{r}^{max}= \max_{\forall t} \;P_{r}(t).
\label{eq:D2}
\end{eqnarray}
Figures \ref{fig10}(a), \ref{fig10}(b) and \ref{fig10}(c) show $P_r(t)$ 
for $d=0.0$, $-0.01$ and $-0.033$, respectively, which are calculated 
by the Gaussian wavepacket with $\mu=0.1$.
For $d=0.0$, $P_r(t)$ oscillates with a period of about 260,
as shown in Figs. \ref{fig4} and \ref{fig5}. 
For $d=-0.01$, $P_r(t)$ almost stay at about 0.2
where it significantly fluctuates.
For $d=-0.033$, $P_r(t)$ again oscillates with a period of
about 60, as shown in Figs. \ref{fig8}(c) and \ref{fig8}(d).

\begin{figure}
\begin{center}
\includegraphics[keepaspectratio=true,width=70mm]{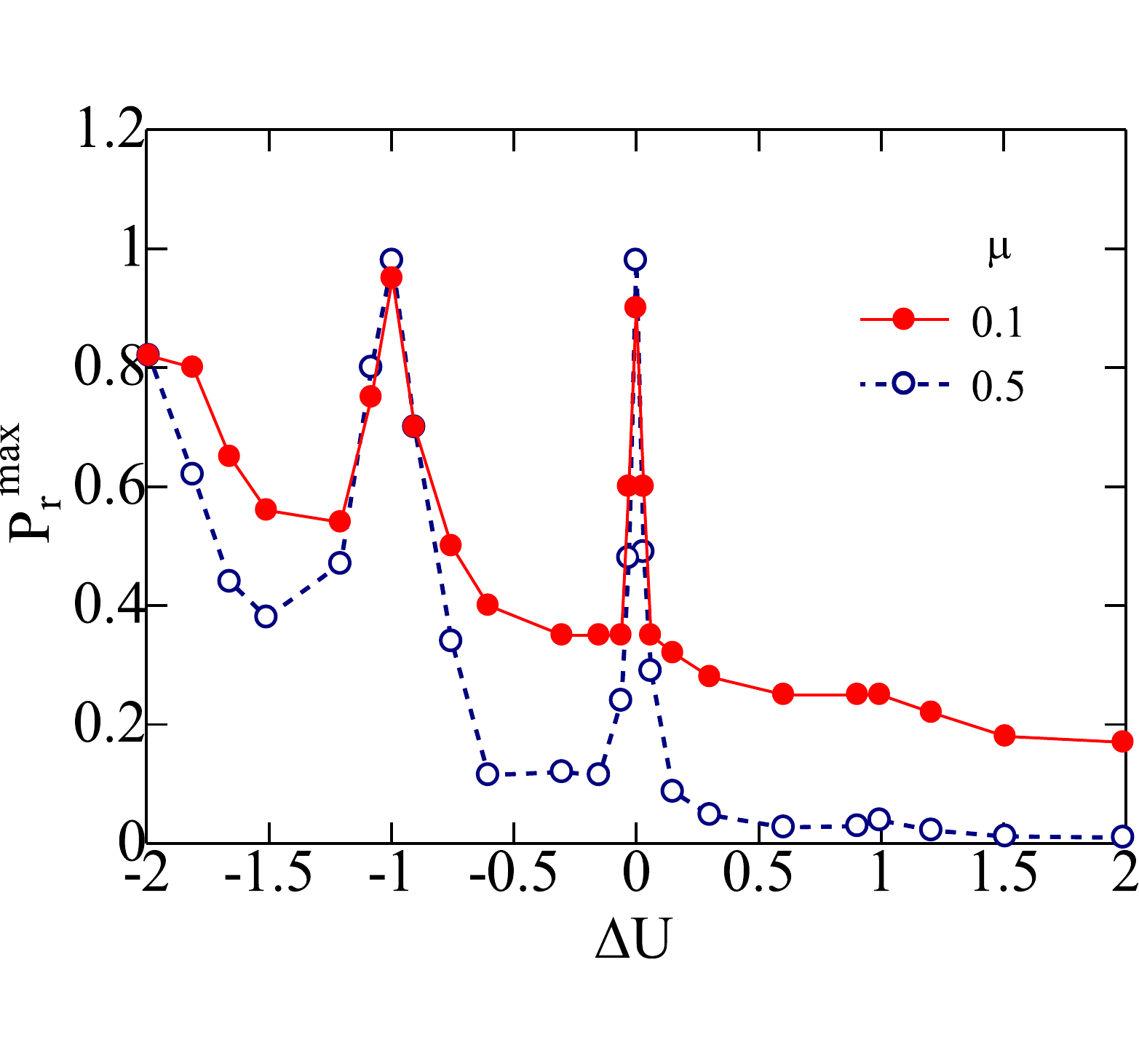}
\end{center}
\caption{
(Color online) 
$P_r^{max}$ as a function of $\Delta U$ calculated by Gaussian wavepackets
with $\mu=0.1$ (solid curve) and $\mu=0.5$ (dashed curve).
}
\label{fig11}
\end{figure}

The maximum value of $P_r^{max}$ is plotted as a function of $\Delta U$ in Fig. \ref{fig11} 
where solid and dashed curves show the results calculated by 
Gaussian wavepackets with $\mu=0.1$ and $\mu=0.5$, respectively.
The maximum value of $P_r^{max} \sim 1.0$ for symmetric DW case ($d=0.0$) is considerably reduced
by an introduced small asymmetry.
For a negative $\Delta U=-0.995$ ($d=-0.033$), $P_r^{max}$ shows an enhanced value
due to a resonance effect, while there is no resonance for a positive $\Delta U=0.995$ ($d=0.033$).
Similarly, the resonant tunneling is realized for a negative $\Delta U=-1.991$ ($d=-0.066$)
but not for a positive $\Delta U=1.991$ ($d=0.066$).
The reduction of $P_r^{max}$ by the asymmetry for the Gaussian wavepacket with $\mu=0.5$
is more significant than that with $\mu=0.1$.
The $\Delta U$ dependence of $P_r^{max}$ is not symmetric with respect to a sign of $\Delta U$,
which is in contrast with the result of Ref. \cite{Mugnai88}.

\section{Discussion}
\subsection{The potential-asymmetry dependence of $P_r^{max}$}
We will discuss the $d$ (or $\Delta U$) dependence of the tunneling probability, 
by using the spectral method B presented in Sec. II B. 
From Eqs. (\ref{eq:C4}) and (\ref{eq:D1}), the tunneling probability is expressed by
\begin{eqnarray}
P_r(t) &=& \sum_{\nu=0}^{N_m} \sum_{\lambda=0}^{N_m} 
a_{\nu}^* a_{\lambda} \;D_{\nu \lambda} \:e^{i \:\Delta E_{\nu \lambda}\:t/\hbar},
\label{eq:D3}
\end{eqnarray}
with
\begin{eqnarray}
D_{\nu \lambda} &=& \int_0^{\infty} \Psi_{\nu}(x)^* \Psi_{\lambda}(x)\:dx,
\label{eq:D4}
\end{eqnarray}
where $\Delta E_{\nu \lambda}= E_{\nu}-E_{\lambda}$.
When main contributions arise from the two terms of $\nu=i$ and $\nu=j$ in Eq. (\ref{eq:C4}),
we may adopt the two-level approximation given by
\begin{eqnarray}
\Psi(x,t) &\simeq& a_{i} \:\Psi_{i}(x)\: e^{- iE_{i} t/\hbar}
+ a_{j} \:\Psi_{j}(x)\: e^{- iE_{j} t/\hbar} 
\hspace{0.5cm}\mbox{$(a_{i}^2+a_{j}^2=1)$},
\label{eq:D6}
\end{eqnarray}
leading to the tunneling probability
\begin{eqnarray}
P_r(t) &\simeq& \vert a_{i} \vert^2 D_{i i}+ \vert a_{j} \vert^2 D_{j j} 
+ 2 \:\Re[a_{i}^* a_{j} D_{i j} \:e^{i \Delta E_{i j}\:t/\hbar }].
\label{eq:D5}
\end{eqnarray}
Equations (\ref{eq:C6}), (\ref{eq:D3}) and (\ref{eq:D4}) signify that
$P_r(t)$ depends on the Gaussian wavepacket $\Psi_G(x,0)$ and
the asymmetry $d$ through the $d$-dependent $a_{\nu}$, $E_{\nu}$ and $\Psi_{\nu}(x)$.

\begin{figure}
\begin{center}
\includegraphics[keepaspectratio=true,width=70mm]{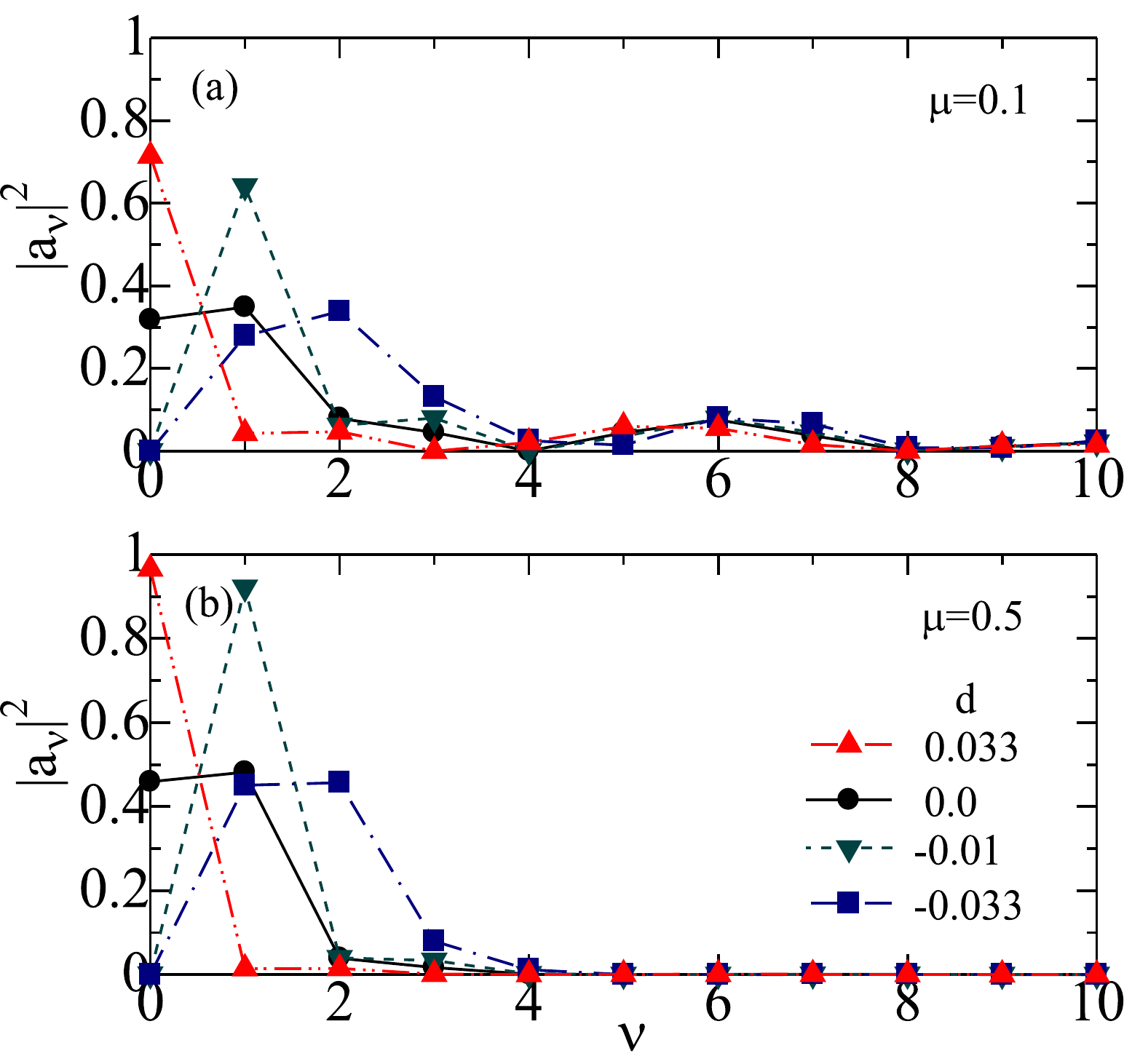}
\end{center}
\caption{
(Color online) 
Magnitudes of expansion coefficients $\vert a_{\nu} \vert^2$ [Eq. (\ref{eq:C4})] against $\nu$ 
with $d=0.033$ (triangles), $d=0$ (circles),
$d=-0.01$ (inverted triangles) and $d=-0.033$ (squares) for Gaussian wavepackets 
with (a) $\mu=0.1$ and (b) $\mu=0.5$.
}
\label{fig12}
\end{figure}

Circles in Figs. \ref{fig12}(a) and \ref{fig12}(b) show magnitudes 
of calculated expansion coefficients $\vert a_{\nu} \vert^2$ 
of symmetric DW systems ($d=0.0$) for Gaussian wavepackets 
with $\mu=0.1$ and $\mu=0.5$, respectively. The magnitude of
$\vert a_{\nu} \vert^2$ for $\mu=0.5$ in Fig. \ref{fig12}(b)
has main contributions from $\nu=0$ and $\nu=1$, which is similar 
to the conventional two-level wavepacket: $\Psi(x,0)=a_0 \Psi_0(x)+a_1 \Psi_1(x)$
with $\vert a_0 \vert^2=\vert a_1 \vert^2= 0.5$
plotted as the dashed curve in Fig. \ref{fig2}.
In contrast, $\vert a_{\nu} \vert^2$ for $\mu=0.1$ in Fig. \ref{fig12}(a)
has extra contributions from $\nu=5-7$ besides those from $\nu=0$ and $\nu=1$
although magnitudes of the former are smaller than those of the latter.
From Eq. (\ref{eq:D5}) the transition probability for $d=0.0$ ($\Delta U=0.0$) is given by
\begin{eqnarray}
P_r(t) &\simeq& a_0^2 \:D_{00} + a_1^2  \:D_{11} 
+ 2 a_0 a_1 \:D_{01} \cos(\Delta E_{01}\:t/\hbar)
\hspace{0.5cm}\mbox{for $\Delta U=0.0$}.
\label{eq:D9}
\end{eqnarray}
Although Eq. (\ref{eq:D9}) leads to a sinusoidal oscillation,
$P_r(t)$ in Fig. \ref{fig10}(a) includes fine structures, which arise
from high-frequency contributions neglected in the two-level approximation
in Eq. (\ref{eq:D6}). 
The essential feature of $P_r(t)$ for $d=0.0$ in Fig. \ref{fig10}(a) may be explained 
by Eq. (\ref{eq:D9}) with $\vert \Delta E_{01} \vert=\delta=0.023923$ 
which yields $T=2 \pi/\delta=262$.

When an asymmetry of $d=-0.01$ ($\Delta U=- 0.3016$) is introduced, the dominant contribution comes 
from $\nu=1$ both for $\mu=0.1$ and $\mu=0.5$ as shown by inverted triangles in
Figs. \ref{fig12}(a) and \ref{fig12}(b).
This suggests that the wavefunction for $d=-0.01$ may be 
given by the one-level state which
yields the time-independent tunneling probability given by
\begin{eqnarray}
P_r(t) &\simeq& a_1^2 \:D_{11}
\hspace{1.0cm}\mbox{for $\Delta U=-0.3016$}.
\label{eq:D7}
\end{eqnarray}
Our calculation of $P_r(t)$ for $\Delta U=- 0.3016$ ($d=-0.01$) in Fig. \ref{fig10}(b) 
shows wiggles, which arise from high-energy contributions
not taken into account in the one-level approximation. 

\begin{figure}
\begin{center}
\includegraphics[keepaspectratio=true,width=70mm]{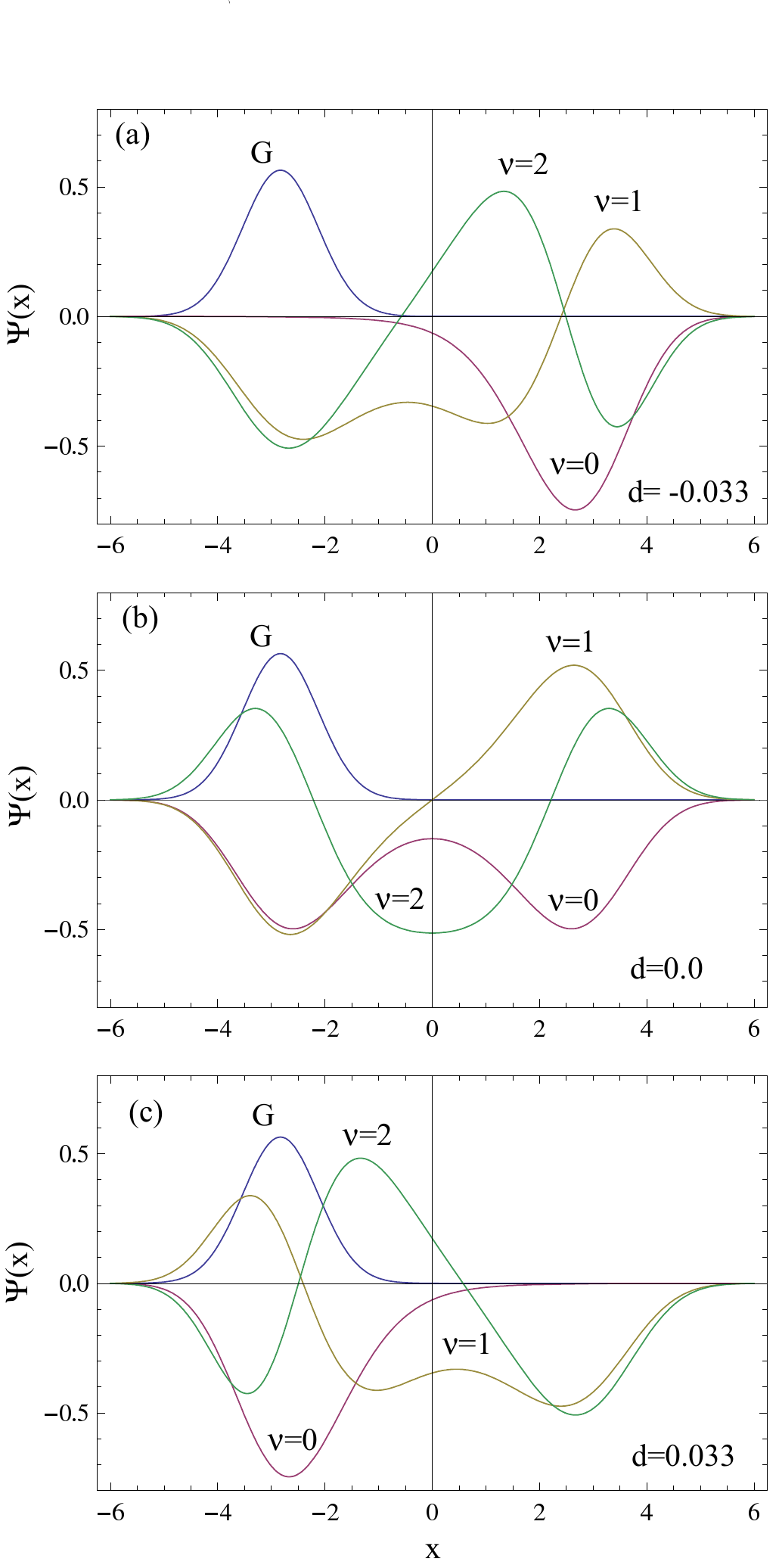}
\end{center}
\caption{
(Color online) 
Gaussian wavepackets of $\Psi_G(x,0)$ (G) with $\mu=0.5$ 
and eigenfunctions of $\Psi_{\nu}(x)$ with
$\nu=0$, 1 and 2 for (a) $d=-0.033$, (b) $d=0.0$ and (c) $d=0.033$.
}
\label{fig13}
\end{figure}

When an asymmetry is increased to $d=-0.033$ ($\Delta U=-0.9956$), 
main contributions to $\vert a_{\nu} \vert^2$ come from $\nu=1$ and $\nu=2$,
as shown by squares in Figs.  \ref{fig12}(a) and \ref{fig12}(b).
From Eq. (\ref{eq:D5}), we obtain the tunneling probability 
\begin{eqnarray}
P_r(t) &\simeq& 
a_1^2 \:D_{11} + a_2^2  \:D_{22} 
+ 2 a_1 a_2 \:D_{12} \cos(\Delta E_{12}\:t/\hbar)
\hspace{0.5cm}\mbox{for $\Delta U=- 0.9956$}.
\label{eq:D10}
\end{eqnarray}
Indeed, $P_r(t)$ for $d=-0.033$ in Fig. \ref{fig10}(c) oscillates 
with a period of about 60 which is consistent with $T=2 \pi/\delta'=59.382$ 
for $\vert \Delta E_{12} \vert =\delta'=0.10581$ (Table 1).

For a negative $\Delta U$, the $\nu=0$ contribution to $\vert a_{\nu} \vert^2$ 
is completely suppressed as shown in Fig. \ref{fig12}.
It is, however, not the case for a positive $\Delta U$ where the $\nu=0$ contribution
is predominant as shown by triangles for $d=0.033$ ($\Delta U =0.996$) 
in Fig. \ref{fig12}.
The wavefunction is approximately expressed by the single $\nu=0$ state
which yields the time-independent tunneling probability
\begin{eqnarray}
P_r(t) &\simeq& a_0^2 \:D_{00}
\hspace{1.0cm}\mbox{for $\Delta U=0.9956$}.
\label{eq:D11}
\end{eqnarray}
The result for a positive $\Delta U=0.9956$ is in contrast to that 
for a negative $\Delta U=-0.9956$ given by Eq. (\ref{eq:D10}).

In the following, we will elucidate
the difference between the $\nu$ dependence of $\vert a_{\nu} \vert^2$ 
for $\Delta U=-0.9956$ and $\Delta U=0.9956$, which
may be understood from Eq. (\ref{eq:C6}) expressed in terms of 
the eigenfunction of $\Psi_{\nu}(x)$ and the Gaussian wavepacket of $\Psi_G(x,0)$.
Eigenfunctions $\Psi_{\nu}(x)$ ($\nu=0$ to $2$) for the asymmetry
of $d=-0.033$, $d=0.0$ and $d=0.033$ are plotted in Figs. \ref{fig13}(a), \ref{fig13}(b) 
and \ref{fig13}(c), respectively, where $\Psi_G(x,0)$ with $\mu=0.5$ is also shown. 
Figure \ref{fig13}(b) shows that the ground-state eigenfunction $\Psi_0(x)$ 
for $d=0.0$ has the equal magnitude at $x= \pm x_s$.
In contrast, $\Psi_0(x)$ for $d=-0.033$ at $x=x_s$
has a larger magnitude than that at $x=-x_s$ as shown in Fig. \ref{fig13}(a).
On the other hand, Fig. \ref{fig13}(c) shows that the situation is reverse
for $d=0.033$: $\vert \Psi_0(-x_s) \vert^2 > \vert \Psi_0(x_s) \vert^2$.
We note in Fig. \ref{fig13}(a) that magnitudes of $a_{1}$ and $a_{2}$ 
for $d=-0.033$ ($\Delta U=-0.9956$) may be appreciable 
because $\Psi_{1}(x)$ and $\Psi_{2}(x)$ overlap with $\Psi_G(x,0)$.
In contrast, Fig. \ref{fig13}(c) shows that
$a_{1}$ and $a_2$ for $d=0.033$ ($\Delta=0.9956$) become 
very small because $\Psi_{1}(x)$ and $\Psi_{2}(x)$ have nodes
near the center of $\Psi_G(x,0)$ while $a_0$ is appreciable because
$\Psi_0(x)$ and $\Psi_G(x,0)$ are overlap.
It is necessary to note that high-energy contributions to $\vert a_{\nu} \vert^2$ 
at $\nu=5-7$ for $\mu=0.1$ are almost independent of the asymmetry in Fig. \ref{fig12}(a)
while there are no such high-energy contributions for $\mu=0.5$ in Fig. \ref{fig12}(b).
This is the reason why the $\Delta U$ dependence of $P_r^{max}$ for $\mu=0.1$
is smaller than that for $\mu=0.5$ as shown in Fig. \ref{fig11}.

When the asymmetry is much increased up to $d=\pm 0.066$ ($\Delta=\pm 1.991$),
the energy gap between the second- and third-excited states:
$E_3-E_2=0.230984$ becomes smaller than $\delta$ and $\delta'$
with $E_2 \;(=1.25498) \lesssim U(x_u)\;(=1.27231) < E_3\;(=1.48597)$
[see Fig. \ref{fig1} (b)]. Then for a negative $\Delta=- 1.991$ ($d=- 0.066$)
dominant contributions to $\vert a_{\nu} \vert^2$ arise from two levels of $\nu=2$ and $\nu=3$, 
while for a positive $\Delta=1.991$ ($d=0.066$) a contribution
from a single level of $\nu=0$ is predominant (relevant results not shown).
This is similar to the case of $\Delta U=-0.9956$ and $\Delta U=0.9956$ mentioned above.
We may similarly elucidate the difference between $P_r^{max}$ 
of $\Delta U=-1.991$ and $\Delta U=1.991$ in Fig. \ref{fig11}.

Cordes and Das (CD) \cite{Cordes01} discussed the tunneling probability in asymmetric DW systems,
proposing the generalized two-level wavefunction given by
\begin{eqnarray}
\Psi^{CD}(x,t)=a_{i} \:\Psi^{CD}_{i}(x)\: e^{- iE_{i} t/\hbar}
+ a_{j} \:\Psi^{CD}_{j}(x)\: e^{- iE_{j} t/\hbar} 
\hspace{1cm}\mbox{$(a_{i}^2+a_{j}^2=1)$}.
\label{eq:K1}
\end{eqnarray}
Here eigenfunctions $\Psi^{CD}_{i}(x)$ and $\Psi^{CD}_{j}(x)$ of the DW system 
are assumed to be expressed by superposition of eigenfunctions for
two harmonic potentials in left and right wells which are separated 
by a high central barrier.
By using Eq. (\ref{eq:K1}), CD showed that the tunneling probability
is given by \cite{Cordes01}
\begin{eqnarray}
P_r^{CD}(t) &=& 2 a_{i}^2 a_{j}^2 [1- \cos(\Delta E_{ij} \:t/\hbar)].
\label{eq:K2}
\end{eqnarray}
The tunneling probability given by Eq. (\ref{eq:K2}) is consistent with our results
for $d=0.0$ and $d=-0.033$ given by Eqs. (\ref{eq:D9}) and (\ref{eq:D10}), respectively.
However Eq. (\ref{eq:K2}) is not valid for cases of $d=-0.01$ and $d=0.033$
for which it yields $P_r^{CD}(t)=0$ in contrast to Eqs. (\ref{eq:D7}) and (\ref{eq:D11}).
Actually, $P_r^{CD}(t)$ in Eq. (\ref{eq:K2}) cannot be applied to
the one-level state with either $a_{i}=0$ or $a_{j}=0$, while
$P_r(t)$ given by Eq. (\ref{eq:D5}) is applicable.

\begin{figure}
\begin{center}
\includegraphics[keepaspectratio=true,width=80mm]{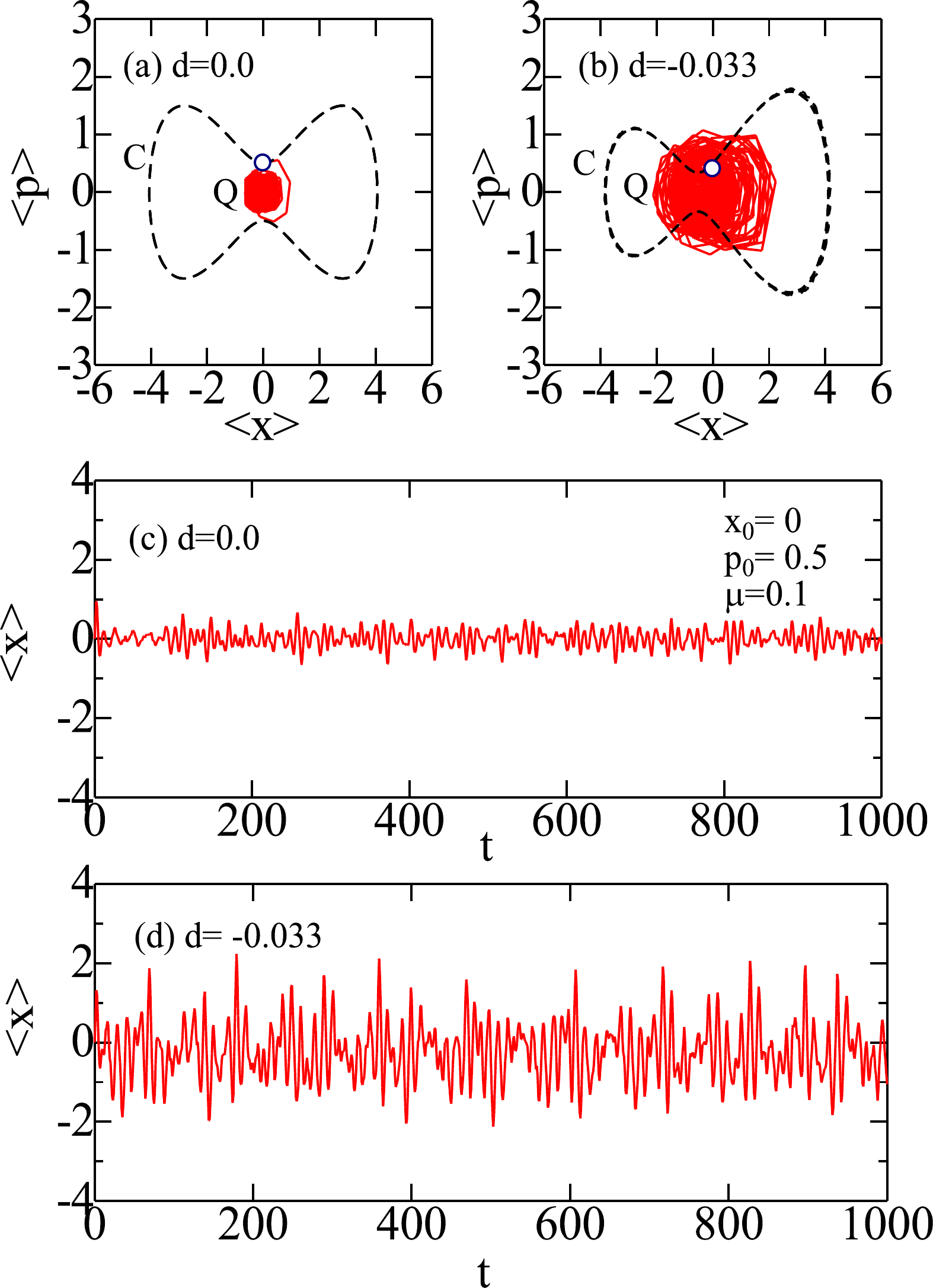}
\end{center}
\caption{
(Color online) 
The $\langle x \rangle$ vs. $\langle p \rangle$ plot for (a) $d=0.0$ and (b) $d=-0.033$
in the classical (C) (dashed curves) and quantum (Q) calculations (solid curves)
with a time step of $\Delta t=1.0$ for $t=0.0 -1000.0$
when a particle starts from the origin near the top of the DW potential,
open circles denoting the starting point of $(x_0, p_0)=(0.0, 0.5)$.  
The time dependence of $\langle x \rangle$ for (c) $d=0.0$ and (d) $d=-0.033$.
Initial Gaussian wavepackets are given by Eq. (\ref{eq:B11})
with $x_0=0.0$, $p_0=0.5$, $\mu=0.1$ and $\alpha=0.0$.
}
\label{fig14}
\end{figure}

\subsection{Varying initial Gaussian wavepacket}
In our study reported in Secs. II and III, we have adopted the initial 
squeezed Gaussian wavepacket given by Eq. (\ref{eq:B11})
with $x_0=-2 \sqrt{2}$, $p_0=0.0$, $\alpha=0.0$ and $\mu=0.1$ (or $\mu=0.5$). 
We may, however, employ any arbitrary initial wavepacket with appropriate parameters
of $x_0$, $p_0$, $\mu$ and $\alpha$, while conventional theories rely on
the two-level wavepacket.
For example, we here employ an initial Gaussian wavepacket with
$x_0=0.0$, $p_0=0.5$, $\mu=0.1$ and $\alpha=0.0$.
In the classical mechanics, a particle with this initial condition rolls down 
from the origin near a top of the potential with an initial velocity of $v_0=p_0/m=0.5$, 
and it continues an oscillation at $-4 \lesssim x \lesssim 4$.
Our calculation, however, shows that motion of a particle in quantum mechanics 
is quite different. The quantum average of $x$ is given by
$-0.5 \lesssim \langle x \rangle \lesssim 0.5$ for $d=0.0$
and $-2 \lesssim \langle x \rangle \lesssim 2$ for $d=-0.033$
as shown in Figs. \ref{fig14}(a)-\ref{fig14}(d). Quantum motion almost stays
near the starting origin: it is difficult for a quantum particle to go across valleys located 
at $x = \pm 2 \sqrt{2}$ [Fig. \ref{fig1}(a)].
The $\langle x \rangle$ vs. $\langle p \rangle$ plot in
Figs. \ref{fig14}(a) and \ref{fig14}(b) shows that
although classical paths are closed in the $\langle x \rangle$-$\langle p \rangle$ space, 
quantum ones are not because of chaotic motion 
which is induced by quantum fluctuations
as pointed out by Pattanayak and Schieve \cite{Pattanayak94}.
We note that results for the initial condition of $(x_0,p_0) =(0.0, 0.5)$
in Figs. \ref{fig14}(c) and \ref{fig14}(d) are quite different 
from relevant results for $(x_0,p_0) =(-2\sqrt{2}, 0.0)$
shown in Figs. \ref{fig5}(a) and \ref{fig8}(c).
Thus the time dependence of $\langle x \rangle$ and other quantities 
depend on the assumed initial condition.

\section{Conclusion}
Dynamics of Gaussian wavepackets and quantum tunneling
in asymmetric DW systems have been studied with the use of the numerical method 
which has advantages that
(a) it is simple and physically transparent,
(b) it is applicable to realistic DW potentials, and
(c) it may adopt an arbitrary, appropriate initial state.
Our calculations have shown the following:

\vspace{0.2cm}
\noindent
(1) The maximum tunneling probability $P_r^{max}$ is considerably reduced 
by a small amount of the asymmetry $\Delta U$ in the DW potential,

\noindent
(2) A resonant tunneling at $\vert \Delta U \vert \simeq \kappa\:\hbar \omega$ 
($\kappa=0, 1, \cdots$) 
is not possible for motion starting from the lower minimum ($\Delta U > 0$) although
it is possible for motion from upper minimum ($\Delta U < 0$) (Fig. 11),

\noindent
(3) $P_r^{max}$ for the Gaussian wavepacket with narrower width ($\mu$) is less
sensitive to the asymmetry, and

\noindent
(4) The uncertainty product in the resonant tunneling state is larger than that 
in the non-resonant tunneling state.

\noindent
The item (1) for $\Delta U \leq 0$ is consistent with results of previous studies
\cite{Weiner81,Nieto85,Cordes01}. 
The item (2) is against Ref. \cite{Mugnai88} which claimed 
the symmetric behavior for motion starting from the upper and lower minima. 
The item (3) is due to the fact that the Gaussian wavepacket with a small $\mu\;(=0.1)$
includes high-energy contributions to $a_{\nu}$ whose magnitudes are nearly 
independent of the asymmetry (Fig. \ref{fig12}).
The item (4) signifies that tunneling and uncertainty, both of which are typical quantum
phenomena, are mutually related.
In order to examine a validity of items (1)-(4),
it would be interesting to observe $\vert \Psi(x,t) \vert^2$ 
in asymmetric DW systems, which seems difficult but possible 
with the recent advance of experimental methods. 
The present study has been made without considering dissipative effects which are expected
to play important roles in stationary and dynamical properties
of real DW systems. An inclusion of dissipation arising from environments 
is left as our future subject.

\begin{acknowledgments}
This work is partly supported by
a Grant-in-Aid for Scientific Research from 
Ministry of Education, Culture, Sports, Science and Technology of Japan.  
\end{acknowledgments}

\appendix*

\section{Matrix elements and various expectation values}
\renewcommand{\theequation}{A\arabic{equation}}
\setcounter{equation}{0}

Matrix elements $H_{nk}$ in Eq. (\ref{eq:B9}) for the adopted DW potential 
are given as follows: We first rewrite $U(x)$ given by Eq. (\ref{eq:L1}) as
\begin{eqnarray}
U(x) &=& \frac{A_4 x^4}{4}+\frac{A_3 x^3}{3}+\frac{A_2 x^2}{2}+A_1 x+A_0,
\label{eq:H4}
\end{eqnarray}
with
\begin{eqnarray}
A_4 &=&\frac{m \omega^2}{2 x_s^2}, \;\;A_3=-d, \;\;A_2=-\frac{m\omega^2}{2},\;\;
A_1=d \:x_s^2,\;\;A_0=\frac{m \omega^2 x_s^2}{8}.
\end{eqnarray}
After some manipulations with the use of relations given by
\begin{eqnarray}
q &=& \sqrt{\frac{g}{2}}(a^{\dagger}+a),\;\;\;
p = i \frac{\hbar}{\sqrt{2 g}} (a^{\dagger}- a),
\hspace{1cm} \mbox{$\left( g=\frac{\hbar}{m \omega} \right)$}
\label{eq:C1a}\\
a^{\dagger} \:\phi_n &=& \sqrt{n+1} \:\phi_{n+1},\;\; 
a \:\phi_n = \sqrt{n} \:\phi_{n-1}, 
\label{eq:C1b}
\end{eqnarray} 
we obtain the symmetric matrix elements $H_{n k}$ for $n \geq k$ given by 
\begin{eqnarray}
H_{nk} &=& \left[ \left( n+1/2\right) \hbar \omega 
+ \frac{3 A_4 g^2}{16}(2 n^2+2 n +1) +\frac{A_2' \:g}{2}(n+1/2)
+ A_0 \right] \:\delta_{n, k} \nonumber \\
&+& \left[A_3 \left(\frac{g}{2}\right)^{3/2} n \sqrt{n} 
+ A_1 \left( \frac{g}{2} \right)^{1/2} \sqrt{n} \right] \delta_{n-1, k} \nonumber \\
&+&  \left[ \frac{A_4 g^2}{8} (n-1)\sqrt{n(n-1)} 
+ \frac{A_2' \:g}{4} \sqrt{n(n-1)} \right] \:\delta_{n-2, k}  \nonumber \\
&+& \frac{A_3}{3} \left(\frac{g}{2}\right)^{3/2} \sqrt{n(n-1)(n-2)}\: \delta_{n-3, k}
\nonumber \\
&+&  \frac{A_4 g^2}{16} \sqrt{n(n-1)(n-2)(n-3)} \:\delta_{n-4, k},
\label{eq:H6}
\end{eqnarray}
where $A_2'=A_2- m \omega^2$ and $g$ ($=\hbar/m \omega$) is unity for $m=\omega=\hbar=1.0$.

In the spectral method A, various time-dependent quantities may be expressed in terms of
$\{ c_n(t) \}$ as follows:
After some manipulations with the use of the relations
Eqs.(\ref{eq:C1a}) and (\ref{eq:C1b}),
the auto-correlation function is given by
\begin{eqnarray}
C(t) &=& \int_{-\infty}^{\infty} \Psi(x,t)^* \Psi(x,0) \:dx, \\
&=& \sum_{n=0}^{N_m} c_n(t)^*\:c_n(0).
\label{eq:C2}
\end{eqnarray}
Expectation values of $x(t)$ and $p(t)$ are expressed by
\begin{eqnarray}
\langle x(t) \rangle &=& \int_{-\infty}^{\infty} \Psi^*(x,t)\: x \:\Psi(x,t)\: dx, \nonumber \\
&=& \sqrt{\frac{g}{2}}
\sum_n \left[\sqrt{n+1}\: c_{n+1}^*(t) c_n(t)+ \sqrt{n}\: c_{n-1}^*(t) c_n(t) \right], 
\label{eq:C3a} \\
\langle p(t) \rangle 
&=& i \sqrt{\frac{\hbar^2}{2 g}}
\sum_n \left[\sqrt{n+1}\: c_{n+1}^*(t) c_n(t) - \sqrt{n}\: c_{n-1}^*(t) c_n(t) \right], \\
\langle x(t)^2 \rangle
&=& \left( \frac{g}{2} \right)
\sum_n [ \sqrt{(n+1)(n+2)} \:c_{n+2}^*(t) c_n(t)+(2 n+1) \:c_{n}^*(t) c_n(t) 
\nonumber \\
&+& \sqrt{n(n-1)} \:c_{n-2}^*(t) c_n(t) ], \\
\langle p(t)^2 \rangle &=& - \left( \frac{\hbar^2}{2 g} \right)
\sum_n [\sqrt{(n+1)(n+2)} \:c_{n+2}^*(t) c_n(t)-(2 n+1) \:c_{n}^*(t) c_n(t) 
\nonumber \\
&+& \sqrt{n(n-1)} \:c_{n-2}^*(t) c_n(t) ], \\
\langle x(t)p(x)+p(t)x(t) \rangle
&=& i \:\hbar \sum_n [ \sqrt{(n+1)(n+2)} \:c_{n+2}^*(t) c_n(t) \nonumber \\
&-& \sqrt{n(n-1)} \:c_{n-2}^*(t) c_n(t) ]. 
\label{eq:C3b}
\end{eqnarray}

On the contrary, in the spectral method B, calculations of time-dependent averages are
more tedious than those in the spectral method A. 
For example, the expectation value of $x(t)$ is given by
\begin{eqnarray}
\langle x(t) \rangle 
&=& \sum_{\nu} \sum_{\lambda}
\:a_{\nu}^* a_{\lambda} \:X_{\nu \lambda} \:e^{i(E_{\nu}-E_{\lambda})t/\hbar},
\end{eqnarray}
where
\begin{eqnarray}
X_{\nu \lambda} &=& \int_{-\infty}^{\infty}
\Psi_{\nu}(x)^* \:x \:\Psi_{\lambda}(x) \:dx.
\end{eqnarray}

\end{document}